\documentstyle[preprint,aps,epsfig,float]{revtex}

\begin{document}
\draft

\title{A New Geometric Probability Technique for an N-dimensional Sphere and Its Applications to Physics}

\author
{
Shu-Ju Tu\thanks{Email: sjtu@physics.purdue.edu}
and Ephraim Fischbach\thanks{Email: ephraim@physics.purdue.edu}
}

\address{Department of Physics, Purdue University, West Lafayette, Indiana 47907-1396 }

\date{\today}

\maketitle
\begin{abstract}
A new formalism is presented for analytically obtaining the probability density
function, \( P_{n}(s) \), for the distance between two random points in an
\( n \)-dimensional sphere of radius \( R \). Our formalism allows \( P_{n}(s) \)
to be calculated for a sphere having an arbitrary density distribution, and
reproduces the well-known results for the case of a sphere with uniform density.
The results find applications in stochastic geometry, probability distribution
theory, astrophysics, nuclear physics, and elementary particle physics.
\end{abstract}


\section{Introduction}

In a recent paper~\cite{Parry_1}, geometric probability techniques were developed
to calculate the probability density functions (PDFs) which describe the probability
density of finding a distance \( s \) separating two points distributed in
a uniform \( 3 \)-dimensional sphere and in a uniform ellipsoid. Our focus
in the present paper will be on the probability density functions \( P_{n}(s) \)
for an \( n \)-dimensional sphere of radius \( R \) characterized by \( x_{1}^{2}+x_{2}^{2}+\cdots +x_{n}^{2}\leq R^{2} \),
where \( x_{1} \), \( x_{2} \), and \( x_{n} \) are the corresponding Cartesian
coordinates. (In the mathematical literature this is sometimes termed an \( n \)-dimensional
ball). As discussed in Refs.~\cite{Parry_1,Kendall,Santalo,Overhauser}, these
results are of interest both as pure mathematics and as tools in mathematical
physics. Specifically, it was demonstrated in Ref.~\cite{Parry_1} that geometric
probability techniques greatly facilitate the calculation of the self-energies
for spherical matter distributions arising from electromagnetic, gravitational,
or weak interactions. The functional form of \( P_{n}(s) \) is known for a
sphere of uniform density, and hence the object of the present paper is to generalize
the results of Refs.~\cite{Kendall,Santalo,Overhauser} to the case of an arbitrarily
non-uniform density distribution by using our method. As an application of these
results we will consider the neutrino-exchange contribution to the self-energy
of a neutron star modeled as series of concentric shells of different constant
density. Other applications will also be discussed.

In this paper we present a new technique for obtaining the analytical probability
density function \( P_{n}(s) \) for a sphere of \( n \) dimensions having
an arbitrary density distribution. To illustrate this technique, we begin by
deriving the PDF, \( P_{n}(s) \), for an \( n \)-dimensional uniform sphere
of radius \( R \), and compare our results to those obtained earlier by other
means~\cite{Kendall,Santalo,Overhauser}. We then extend this technique to an
\( n \)-dimensional sphere with a non-uniform but spherically symmetric density
distribution. We explicitly evaluate the analytical probability density functions
for certain specific density distributions, and then use numerical Monte Carlo
simulations to verify the analytical results. 

Finally our formalism is generalized to an \( n \)-dimensional sphere with
an arbitrary density distribution, and leads to a general-purpose master formula.
This formula allows one to evaluate the PDF for a sphere in \( n \) dimensions
with an arbitrary density distribution. After verifying that the master formula
reproduces the results for uniform and spherically symmetric density distributions,
we analytically evaluate the probability density functions, for 2, 3, and 4-dimensional
spheres having non-uniform density distributions. The analytical results are
then verified by the use of Monte Carlo simulations. The outline of this paper
is as follows. In Sec.~\ref{sec_uniform_density} we present our new formalism
and illustrate it by rederiving the well-known results for a circle and for
a sphere of uniform density. In Sec.~\ref{sec_spherical_density} we extend
this formalism to the case of non-uniform but spherically symmetric density
distributions. In Sec.~\ref{sec_arbitrary_density} we develop the formalism
for the most general case of an arbitrary non-uniform density distribution.
In Sec.~\ref{applications} we present some applications of our formalism to
physics. These include the \( m \)th moment \( \left\langle s^{m}\right\rangle  \)
for a sphere of uniform and Gaussian density distribution, Coulomb self-energy
for a collection of charges, \( \nu \bar{\nu } \)-exchange interactions, obtaining
the probability density functions for multiple-shell density distributions found
in neutron star models~\cite{Pines,Shapiro}, and the evaluation of some geometric
probability constants~\cite{sjtu,sjtu2}.

\section{Uniform density distributions}
\label{sec_uniform_density}

\subsection{Theory for a uniform circle}
\label{section_theory_uniform_circle}

\begin{figure}
\phantom{aaa}
\vfill
{\par\centering \resizebox*{6.25in}{!}{\includegraphics{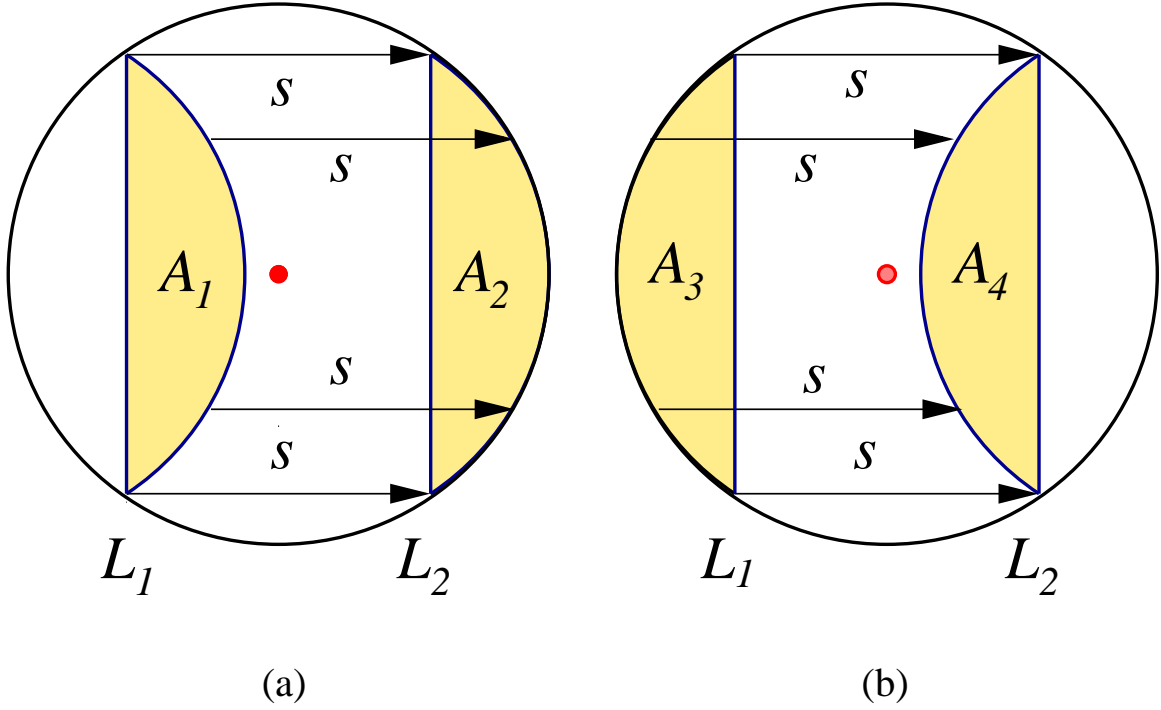}} \par}
\caption{(a) Locus of points in a circle of radius \protect\( R\protect \) separated
by a vector \protect\( \vec{s}=s\hat{x}\protect \). For each point \protect\( \vec{r}_{1}\protect \)
in \protect\( A_{1}\protect \), there is a unique point \protect\( \vec{r}_{2}\protect \)
in \protect\( A_{2}\protect \) such that \protect\( \vec{s}=\vec{r}_{2}-\vec{r}_{1}\protect \).
\protect\( A_{2}\protect \) is the overlap between the circle and the region
obtained by translating the line \protect\( L_{2}\protect \) given by \protect\( x=s/2\protect \)
to the \protect\( +\hat{x}\protect \) direction. \protect\( A_{1}\protect \)
can be obtained by shifting \protect\( A_{2}\protect \) a distance \protect\( s\protect \)
along the \protect\( +\hat{x}\protect \) direction. (b) \protect\( A_{3}\protect \)
is the overlap between the circle and the region obtained by translating the
line \protect\( L_{1}\protect \) given by \protect\( x=-s/2\protect \) along
the \protect\( -\hat{x}\protect \) direction. \protect\( A_{4}\protect \)
can be obtained by shifting \protect\( A_{3}\protect \) a distance \protect\( s\protect \)
along the \protect\( \hat{x}\protect \) direction. \label{fig_choice_2}}
\vfill\phantom{aaa}\end{figure}

In this section we illustrate our formalism by deriving the PDF for a circle
of radius \( R \) having a spatially uniform density distribution characterized
by \( \rho = \) constant. For two points randomly sampled inside the circle
located at \( \vec{r}_{1} \) and \( \vec{r}_{2} \) measured from the center,
define \( \vec{s}=\vec{s}\left( \vec{r}_{1},\vec{r}_{2}\right) =\vec{r}_{2}-\vec{r}_{1} \)
and \( s=\left| \vec{s}\, \right|  \). To simplify the discussion, we translate
the center of the circle to the origin so that the equation for the circle is
\( x^{2}+y^{2}=R^{2} \). It is sufficient to initially consider those vectors
\( \vec{s} \) which are aligned in the positive \( \hat{x} \) direction, since
rotational symmetry can eventually be used to extend our results to those vectors
\( \vec{s} \) with arbitrary orientations. We begin by identifying those pairs
of points, \( \vec{r}_{1} \) and \( \vec{r}_{2} \), which satisfy \( \vec{s}=\vec{r}_{2}-\vec{r}_{1}=s\hat{x} \),
where \( 0\leq s\leq 2R \). One set of points is obtained if the points 1 are
uniformly located on a line \( L_{1} \) given by \( x=-s/2 \) and the corresponding
points 2 are on a line \( L_{2} \) given by \( x=+s/2 \) as shown in Fig.~\ref{fig_choice_1}.
Notice that \( x=\pm s/2 \) are two symmetric lines with respect to reflection
about \( x=0 \). Another set of points is obtained if \( \vec{r}_{1} \) is
uniformly located in the area \( A_{1} \) and \( \vec{r}_{2} \) is located
in the area \( A_{2} \) as shown in Fig.~\ref{fig_choice_2}(a). Note that
\( A_{1} \) and \( A_{2} \) are congruent. The only remaining set of points
that need to be considered are those for which \( \vec{r}_{1} \) is uniformly
located in the area \( A_{3} \) while \( \vec{r}_{2} \) is in the area \( A_{4} \)
as shown in Fig.~\ref{fig_choice_2}(b). As before \( A_{3} \) and \( A_{4} \)
are congruent. Furthermore, \( A_{4} \) is a reflection of \( A_{2} \) about
\( x=s/2 \), while \( A_{3} \) is a reflection of \( A_{1} \) about \( x=-s/2 \).
As discussed in Appendix~\ref{appendix_geometry}, the union of \( A_{2} \)
and \( A_{4} \) (\( A_{2}\bigcup A_{4} \)) is the overlap area between the
original circle \( C_{0} \) and an identical circle \( C_{1} \) whose center
is shifted from \( \left( 0,0\right)  \) to \( \left( \left| \vec{s}\, \right| ,0\right)  \).
Similarly, \( A_{1}\bigcup A_{3} \) is the overlap area between the original
circle and an identical circle \( C_{2} \) whose center is shifted from \( \left( 0,0\right)  \)
to \( \left( -\left| \vec{s}\, \right| ,0\right)  \) as shown in Fig.~\ref{fig_shift_center}.
Since \( A_{1}\bigcup A_{3} \) and \( A_{2}\bigcup A_{4} \) are identical,
it follows that the probability of finding a given \( s=\left| s\hat{x}\right|  \)
in a uniform circle is proportional to \begin{equation}
A_{2}\bigcup A_{4}=\int _{\frac{s}{2}}^{R}dx\int _{-\sqrt{R^{2}-x^{2}}}^{\sqrt{R^{2}-x^{2}}}dy+\int _{s-R}^{\frac{s}{2}}dx\int _{-\sqrt{R^{2}-(x-s)^{2}}}^{\sqrt{R^{2}-(x-s)^{2}}}dy.
\end{equation}
 Notice that \[
\int _{\frac{s}{2}}^{R}dx\int _{-\sqrt{R^{2}-x^{2}}}^{\sqrt{R^{2}-x^{2}}}dy=\int _{s-R}^{\frac{s}{2}}dx\int _{-\sqrt{R^{2}-(x-s)^{2}}}^{\sqrt{R^{2}-(x-s)^{2}}}dy.\]
Using rotational symmetry of \( C_{0} \), this result can apply to any orientation
of \( \vec{s} \) between \( 0 \) and \( 2\pi  \), and hence the probability
of finding a given \( s=\left| \vec{s}\, \right|  \) is proportional to \( |\vec{s}\, |\int _{0}^{2\pi }d\phi =2\pi s \).
We denote this PDF for a uniform circle by \( P_{2}(s) \), and impose the normalization
requirement \begin{equation}
\label{2_002}
\int _{0}^{2R}P_{2}(s)ds=1.
\end{equation}
 We then have 

\begin{eqnarray}
P_{2}(s) & = & \frac{2\pi s\left( \int _{\frac{s}{2}}^{R}dx\int _{-\sqrt{R^{2}-x^{2}}}^{\sqrt{R^{2}-x^{2}}}dy+\int _{s-R}^{\frac{s}{2}}dx\int _{-\sqrt{R^{2}-(x-s)^{2}}}^{\sqrt{R^{2}-(x-s)^{2}}}dy\right) }{\int _{0}^{2R}2\pi s\left( \int _{\frac{s}{2}}^{R}dx\int _{-\sqrt{R^{2}-x^{2}}}^{\sqrt{R^{2}-x^{2}}}dy+\int _{s-R}^{\frac{s}{2}}dx\int _{-\sqrt{R^{2}-(x-s)^{2}}}^{\sqrt{R^{2}-(x-s)^{2}}}dy\right) ds}.\label{eq_2_001} 
\end{eqnarray}
Equation (\ref{eq_2_001}) can be simplified to \begin{eqnarray}
P_{2}(s) & = & \frac{s\int _{\frac{s}{2}}^{R}dx\int _{-\sqrt{R^{2}-x^{2}}}^{\sqrt{R^{2}-x^{2}}}dy}{\int _{0}^{2R}\left( s\int _{\frac{s}{2}}^{R}dx\int _{-\sqrt{R^{2}-x^{2}}}^{\sqrt{R^{2}-x^{2}}}dy\right) ds}\label{eq_2_002_a} \\
 & = & \frac{2s}{R^{2}}-\frac{s^{2}}{\pi R^{4}}\sqrt{4R^{2}-s^{2}}-\frac{4s}{\pi R^{2}}\sin ^{-1}\left( \frac{s}{2R}\right) .\label{eq_2_002_b} 
\end{eqnarray}
 Equation~(\ref{eq_2_002_b}) is identical to the results obtained in Refs.~\cite{Kendall,Santalo}
by other means.

The conclusion that emerges from this formalism is that the probability of finding
two random points separated by a vector \( \vec{s} \) in a uniform circle can
be derived by simply calculating the overlap region of that circle with an identical
circle obtained by shifting the center from the origin to \( \vec{s} \). In
the following sections, we show that this result generalizes to higher dimensions,
and provides a simple way of calculating \( P_{n}(s) \) for \( n\geq 3 \).

\subsection{Theory for a uniform sphere}

The above formalism for a \( 2 \)-dimensional uniform circle can be extended
to a \( 3 \)-dimensional uniform sphere with radius \( R \). For a given \( s \),
we arbitrarily select the positive \( \hat{z} \) direction and study the distribution
of vectors \( \vec{s}=\vec{r}_{2}-\vec{r}_{1} \) in this direction. The areas
\( A_{1} \), \( A_{2} \), \( A_{3} \), and \( A_{4} \) in Sec.~\ref{section_theory_uniform_circle}
are replaced by the volumes \( V_{1} \), \( V_{2} \), \( V_{3} \), and \( V_{4} \).
The PDF \( P_{3}(s) \) is therefore proportional to the overlap volume \( V_{1}\bigcup V_{3}=V_{2}\bigcup V_{4} \),
where \begin{equation}
V_{2}\bigcup V_{4}=\int _{\frac{s}{2}}^{R}dz\int _{-\sqrt{R^{2}-z^{2}}}^{\sqrt{R^{2}-z^{2}}}dx\int ^{\sqrt{R^{2}-z^{2}-x^{2}}}_{-\sqrt{R^{2}-z^{2}-x^{2}}}dy+\int _{s-R}^{\frac{s}{2}}dz\int _{-\sqrt{R^{2}-(z-s)^{2}}}^{\sqrt{R^{2}-(z-s)^{2}}}dx\int ^{\sqrt{R^{2}-(z-s)^{2}-x^{2}}}_{-\sqrt{R^{2}-(z-s)^{2}-x^{2}}}dy.
\end{equation}
 Notice that \begin{equation}
\int _{\frac{s}{2}}^{R}dz\int _{-\sqrt{R^{2}-z^{2}}}^{\sqrt{R^{2}-z^{2}}}dx\int ^{\sqrt{R^{2}-z^{2}-x^{2}}}_{-\sqrt{R^{2}-z^{2}-x^{2}}}dy=\int _{s-R}^{\frac{s}{2}}dz\int _{-\sqrt{R^{2}-(z-s)^{2}}}^{\sqrt{R^{2}-(z-s)^{2}}}dx\int ^{\sqrt{R^{2}-(z-s)^{2}-x^{2}}}_{-\sqrt{R^{2}-(z-s)^{2}-x^{2}}}dy.
\end{equation}
In a \( 3 \)-dimensional space, \( \vec{s} \) can range from \( 0 \) to \( \pi  \)
in the \( \hat{\theta } \) direction, and from \( 0 \) to \( 2\pi  \) in
the \( \hat{\phi } \) direction. For a given \( s \) with any orientation,
the PDF \( P_{3}(s) \) is also proportional to \( 4\pi s^{2} \) by using rotational
symmetry of a uniform sphere. Following the previous discussion, we thus arrive
at the following expression for the PDF for a uniform sphere: \begin{eqnarray}
P_{3}(s) & = & \frac{4\pi s^{2}\left( \int _{\frac{s}{2}}^{R}dz\int _{-\sqrt{R^{2}-z^{2}}}^{\sqrt{R^{2}-z^{2}}}dx\int ^{\sqrt{R^{2}-z^{2}-x^{2}}}_{-\sqrt{R^{2}-z^{2}-x^{2}}}dy\right) }{\int _{0}^{2R}4\pi s^{2}\left( \int _{\frac{s}{2}}^{R}dz\int _{-\sqrt{R^{2}-z^{2}}}^{\sqrt{R^{2}-z^{2}}}dx\int ^{\sqrt{R^{2}-z^{2}-x^{2}}}_{-\sqrt{R^{2}-z^{2}-x^{2}}}dy\right) ds}\label{eq_2_003_a} \\
 & = & 3\frac{s^{2}}{R^{3}}-\frac{9}{4}\frac{s^{3}}{R^{4}}+\frac{3}{16}\frac{s^{5}}{R^{6}}.\label{eq_2_003_b} 
\end{eqnarray}
 The result in Eq.~(\ref{eq_2_003_b}) agrees exactly with the expression obtained
previously in Refs.~\cite{Kendall,Santalo,Overhauser}.

\subsection{Representations , general properties, recursion relations, and generating functions
of \protect\( P_{n}(s)\protect \)}

The present formalism can be readily generalized to obtain \( P_{n}(s) \) for
an \( n \)-dimensional sphere (\( n \)-sphere or \( n \)-ball) of radius
\( R \). From Eqs.~(\ref{eq_2_002_a}) and (\ref{eq_2_003_a}) we find\begin{equation}
\label{eq_2_100}
P_{n}(s)=\frac{s^{n-1}\int _{\frac{s}{2}}^{R}dx_{n}\int _{-\sqrt{R^{2}-x_{n}^{2}}}^{\sqrt{R^{2}-x_{n}^{2}}}dx_{1}\int _{-\sqrt{R^{2}-x_{n}^{2}-x_{1}^{2}}}^{\sqrt{R^{2}-x_{n}^{2}-x_{1}^{2}}}dx_{2}\cdots \int _{-\sqrt{R^{2}-x_{n}^{2}-\cdots x_{n-2}^{2}}}^{\sqrt{R^{2}-x_{n}^{2}-\cdots x_{n-2}^{2}}}dx_{n-1}}{\int _{0}^{2R}\left( s^{n-1}\int _{\frac{s}{2}}^{R}dx_{n}\int _{-\sqrt{R^{2}-x_{n}^{2}}}^{\sqrt{R^{2}-x_{n}^{2}}}dx_{1}\int _{-\sqrt{R^{2}-x_{n}^{2}-x_{1}^{2}}}^{\sqrt{R^{2}-x_{n}^{2}-x_{1}^{2}}}dx_{2}\cdots \int _{-\sqrt{R^{2}-x_{n}^{2}-\cdots x_{n-2}^{2}}}^{\sqrt{R^{2}-x_{n}^{2}-\cdots x_{n-2}^{2}}}dx_{n-1}\right) ds}.
\end{equation}
 We can rewrite Eq.~(\ref{eq_2_100}) as \begin{equation}
\label{eq_2_112}
P_{n}(s)=\frac{s^{n-1}\int _{\frac{s}{2}}^{R}dx_{n}\int _{-\overline{R}}^{\overline{R}}dx_{1}\int ^{\sqrt{\overline{R}^{2}-x_{1}^{2}}}_{-\sqrt{\overline{R}^{2}-x_{1}^{2}}}dx_{2}\cdots \int _{-\sqrt{\overline{R}^{2}-x_{1}^{2}-\cdots x_{n-2}^{2}}}^{\sqrt{\overline{R}^{2}-x_{1}^{2}-\cdots x_{n-2}^{2}}}dx_{n-1}}{\int _{0}^{2R}\left( s^{n-1}\int _{\frac{s}{2}}^{R}dx_{n}\int _{-\overline{R}}^{\overline{R}}dx_{1}\int ^{\sqrt{\overline{R}^{2}-x_{1}^{2}}}_{-\sqrt{\overline{R}^{2}-x_{1}^{2}}}dx_{2}\cdots \int _{-\sqrt{\overline{R}^{2}-x_{1}^{2}-\cdots x_{n-2}^{2}}}^{\sqrt{\overline{R}^{2}-x_{1}^{2}-\cdots x_{n-2}^{2}}}dx_{n-1}\right) ds},
\end{equation}
 where 

\begin{equation}
\overline{R}=\sqrt{R^{2}-x_{n}^{2}}.
\end{equation}
 As shown in Refs.~\cite{apostol,sommerville,Weisstein}, the volume \( V(n,R) \)
of an \( n \)-dimensional sphere of radius \( R \) is given by \begin{equation}
V\left( n,R\right) =\int _{-R}^{R}dx_{1}\int ^{\sqrt{R^{2}-x_{1}^{2}}}_{-\sqrt{R^{2}-x_{1}^{2}}}dx_{2}\cdots \int _{-\sqrt{R^{2}-x_{1}^{2}-x_{2}^{2}-\cdots x_{n-1}^{2}}}^{\sqrt{R^{2}-x_{1}^{2}-x_{2}^{2}-\cdots x_{n-1}^{2}}}dx_{n}=\frac{\pi ^{\frac{n}{2}}R^{n}}{\Gamma \left( 1+\frac{n}{2}\right) }.
\end{equation}
Equations.~(\ref{eq_2_100}) and (\ref{eq_2_112}) can then be reduced to a
general simplified equation for an \( n \)-dimensional uniform sphere of radius
\( R \): \begin{eqnarray}
P_{n}(s) & = & \frac{s^{n-1}\int _{\frac{s}{2}}^{R}\frac{\pi ^{\frac{n-1}{2}}}{\Gamma \left( 1+\frac{n-1}{2}\right) }\left( \sqrt{R^{2}-x_{n}^{2}}\right) ^{\left( n-1\right) }dx_{n}}{\int _{0}^{2R}\left[ s^{n-1}\int _{\frac{s}{2}}^{R}\frac{\pi ^{\frac{n-1}{2}}}{\Gamma \left( 1+\frac{n-1}{2}\right) }\left( \sqrt{R^{2}-x_{n}^{2}}\right) ^{\left( n-1\right) }dx_{n}\right] ds}\nonumber \\
 & = & \frac{s^{n-1}\int _{\frac{s}{2}}^{R}\left( R^{2}-x^{2}\right) ^{\frac{n-1}{2}}dx}{\int _{0}^{2R}\left[ s^{n-1}\int _{\frac{s}{2}}^{R}\left( R^{2}-x^{2}\right) ^{\frac{n-1}{2}}dx\right] ds}.\label{eq_tu_n_uniform} 
\end{eqnarray}
 If \( n \) is an even number,\begin{equation}
\label{eq_tu_even_uniform}
P_{n}(s)=n\times \frac{s^{n-1}}{R^{n}}\left[ \frac{2}{\pi }\cos ^{-1}\left( \frac{s}{2R}\right) -\frac{s}{\pi }\sum _{k=1}^{\frac{n}{2}}\frac{(n-2k)!!}{(n-2k+1)!!}\left( R^{2}-\frac{s^{2}}{4}\right) ^{\frac{n-2k+1}{2}}R^{2k-2-n}\right] ,
\end{equation}
 where \( 0!=0!!=1 \). If \( n \) is an odd number, \begin{equation}
\label{eq_tu_odd_uniform}
P_{n}(s)=n\times \frac{s^{n-1}}{R^{n}}\frac{n!!}{(n-1)!!}\sum _{k=0}^{\frac{n-1}{2}}\frac{(-1)^{k}}{2k+1}\frac{\left( \frac{n-1}{2}\right) !}{k!\left( \frac{n-1}{2}-k\right) !}\left[ 1-\left( \frac{s}{2R}\right) ^{2k+1}\right] .
\end{equation}
 Using Eqs.~(\ref{eq_tu_even_uniform}) and (\ref{eq_tu_odd_uniform}) the explicit
functional forms of \( P_{n}(s) \) for \( n=2 \), \( 3 \), \( 4 \), and
\( 5 \) are as follows:\begin{eqnarray}
P_{2}(s) & = & \frac{4}{\pi }\frac{s}{R^{2}}\cos ^{-1}\left( \frac{s}{2R}\right) -\frac{2}{\pi }\frac{s^{2}}{R^{3}}\left( 1-\frac{s^{2}}{4R^{2}}\right) ^{1/2},\\
P_{4}(s) & = & \frac{8}{\pi }\frac{s^{3}}{R^{4}}\cos ^{-1}\left( \frac{s}{2R}\right) -\frac{8}{3\pi }\frac{s^{4}}{R^{5}}\left( 1-\frac{s^{2}}{4R^{2}}\right) ^{3/2}-\frac{4}{\pi }\frac{s^{4}}{R^{5}}\left( 1-\frac{s^{2}}{4R^{2}}\right) ^{1/2},\\
P_{3}(s) & = & 3\frac{s^{2}}{R^{3}}-\frac{9}{4}\frac{s^{3}}{R^{4}}+\frac{3}{16}\frac{s^{5}}{R^{6}},\\
P_{5}(s) & = & 5\frac{s^{4}}{R^{5}}-\frac{75}{16}\frac{s^{5}}{R^{6}}+\frac{25}{32}\frac{s^{7}}{R^{8}}-\frac{15}{256}\frac{s^{9}}{R^{10}}.
\end{eqnarray}
 The Monte Carlo results for \( n\geq 4 \), and the simulation techniques for
producing random points uniformly inside an \( n \)-sphere will be presented
elsewhere~\cite{sjtu,sjtu2}. 

It is of interest to verify that Eq.~(\ref{eq_tu_n_uniform}), obtained via
the present formalism, agrees with results obtained earlier by other means~\cite{Kendall,Santalo}.
This is most easily done by introducing the the function \( C(a;\, m,n) \)
defined by \begin{equation}
\label{eq_notation_001}
C(a;\, m,n)=\int _{0}^{a}s^{m}T_{n}(s)ds=\int _{0}^{a}s^{m+n-1}Q_{n}(s)ds=\int _{0}^{a}s^{m+n-1}\, ds\int _{\frac{s}{2}}^{R}\left( R^{2}-x^{2}\right) ^{\frac{n-1}{2}}\, dx,
\end{equation}
where \begin{eqnarray}
Q_{n}(s) & = & \int _{\frac{s}{2}}^{R}\left( R^{2}-x^{2}\right) ^{\frac{n-1}{2}}dx,\label{eq_Q_n} \\
T_{n}(s) & = & s^{n-1}\int _{\frac{s}{2}}^{R}\left( R^{2}-x^{2}\right) ^{\frac{n-1}{2}}dx.\label{eq_T_n} 
\end{eqnarray}
It follows that the denominator of Eq.~(\ref{eq_tu_n_uniform}), which is the
normalization constant, can be written as \( C(2R;\, 0,n) \) where \begin{eqnarray}
C(2R;\, 0,n) & = & \int _{0}^{2R}T_{n}(s)ds.\nonumber 
\end{eqnarray}
 When \( n \) is an even integer, \( C(2R;\, 0,n) \) can be expressed in terms
of the gamma and beta functions by noting that \begin{equation}
C(2R;\, 0,n)=\frac{\pi }{2n}\frac{(n-1)!!}{n!!}R^{2n}=\frac{1}{2n}\frac{\Gamma \left( \frac{1}{2}\right) \Gamma \left( \frac{n}{2}+\frac{1}{2}\right) }{\Gamma \left( \frac{n}{2}+1\right) }R^{2n}=\frac{1}{2n}B\left( \frac{n}{2}+\frac{1}{2},\frac{1}{2}\right) R^{2n}.
\end{equation}
Similarly, when \( n \) is an odd integer, \begin{equation}
C(2R;\, 0,n)=\frac{1}{n}\frac{(n-1)!!}{n!!}R^{2n}=\frac{1}{2n}B\left( \frac{n}{2}+\frac{1}{2},\frac{1}{2}\right) R^{2n}.
\end{equation}
 We thus find that the normalization constant has the same functional form irrespective
of whether \( n \) is even or odd, when expressed in terms of the beta function.

If we introduce the variable \( t=R^{2}-x^{2} \) and note that\begin{equation}
Q_{n}(s)=\frac{R^{n}}{2}B_{1-\frac{s^{2}}{4R^{2}}}\left( \frac{n}{2}+\frac{1}{2},\frac{1}{2}\right) =\frac{R^{n}}{2}B\left( \frac{n}{2}+\frac{1}{2},\frac{1}{2}\right) I_{1-\frac{s^{2}}{4R^{2}}}\left( \frac{n}{2}+\frac{1}{2},\frac{1}{2}\right) ,
\end{equation}
 we can then rewrite \( P_{n}(s) \) in the form \begin{eqnarray}
P_{n}(s) & = & n\frac{s^{n-1}}{R^{n}}\frac{B_{x}\left( \frac{n}{2}+\frac{1}{2},\frac{1}{2}\right) }{B\left( \frac{n}{2}+\frac{1}{2},\frac{1}{2}\right) }\label{eq_tu_001} \\
 & = & n\frac{s^{n-1}}{R^{n}}I_{x}\left( \frac{n}{2}+\frac{1}{2},\frac{1}{2}\right) ,\label{eq_kendall_santalo} 
\end{eqnarray}
 where\begin{equation}
x=1-\frac{s^{2}}{4R^{2}}.
\end{equation}
\( B_{x}\left( p,q\right)  \) is the incomplete beta function defined by \begin{equation}
B_{x}\left( p,q\right) =\int _{0}^{x}t^{p-1}\left( 1-t\right) ^{q-1}dt,
\end{equation}
 and \( I_{x} \) is the normalized incomplete beta function defined by \begin{equation}
I_{x}(p,q)=\frac{\Gamma \left( p+q\right) }{\Gamma \left( p\right) \Gamma \left( q\right) }\int _{0}^{x}t^{p-1}\left( 1-t\right) ^{q-1}dt=\frac{B_{x}(p,q)}{B(p,q)}.
\end{equation}
 Eq.~(\ref{eq_kendall_santalo}) is the expression for \( P_{n}(s) \) obtained
in Refs.~\cite{Kendall,Santalo}, and hence we have demonstrated that the classical
results can be reproduced by using the formalism developed here.

We find the following general properties of \( P_{n}(s) \) and its derivative
\( P'_{n}(s) \) at the lower bound \( s=0 \) and at the upper bound \( s=2R \):
\( P_{1}(0)=1/R \), \( P_{n}(0)=0 \) for \( n\geq 2 \), \( P_{n}(2R)=0 \),
\( P_{1}'(0)=-1/2R^{2} \), \( P_{2}'(0)=2/R^{2} \), \( P_{n}'(0)=0 \) for
\( n\geq 3 \), \( P_{1}'(2R)=-1/2R^{2} \), and \( P_{n}'(2R)=0 \) for \( n\geq 2 \).

It is of interest to express the probability density functions \( P_{n}(s) \)
in terms of generating functions from which several representations and recursion
relations for \( P_{n}(s) \) can be derived. This can be achieved by observing
that \begin{equation}
B_{x}\left( \frac{n}{2}+\frac{1}{2},\frac{1}{2}\right) =\frac{1}{n!}\left( \frac{\partial }{\partial h}\right) ^{n}F(h=0,x),
\end{equation}
where \begin{equation}
\label{eq_generating_F}
F(h,x)=\frac{2}{\sqrt{1-h^{2}}}\left[ \sin ^{-1}(h)\, -\sin ^{-1}\left( \frac{h-\sqrt{x}}{1-h\sqrt{x}}\right) \right] ,
\end{equation}
 and where \( 0\leq x\leq 1 \) and \( -1\leq h\leq 1 \). Similarly, \( Q_{n}(s) \)
in Eq.~(\ref{eq_Q_n}) can be defined in terms of a generating function \( F_{1}(h,s) \)
of the form

\begin{equation}
\label{eq_generating_F1}
F_{1}(h,s)=\sum _{n=0}^{\infty }Q_{n}(s)h^{n}=\frac{1}{\sqrt{1-h^{2}R^{2}}}\left[ \sin ^{-1}(hR)-\sin ^{-1}\left( \frac{hR-\sqrt{1-\frac{s^{2}}{4R^{2}}}}{1-hR\sqrt{1-\frac{s^{2}}{4R^{2}}}}\right) \right] ,
\end{equation}
where \( \left| hR\right| <1 \). \( Q_{n}(s) \) is then given by \begin{equation}
Q_{n}(s)=\frac{1}{n!}\left( \frac{\partial }{\partial h}\right) ^{n}F_{1}(h=0,s).
\end{equation}
 \( T_{n}(s) \) in Eq.~(\ref{eq_T_n}) can also be defined in terms of a generating
function \( F_{2}(h,s) \), \begin{equation}
\label{eq_generating_F2}
F_{2}(h,s)=\sum _{n=0}^{\infty }T_{n}(s)h^{n}=\left( \frac{1}{s}\right) \frac{1}{\sqrt{1-h^{2}R^{2}s^{2}}}\left[ \sin ^{-1}(hRs)-\sin ^{-1}\left( \frac{hRs-\sqrt{1-\frac{s^{2}}{4R^{2}}}}{1-hRs\sqrt{1-\frac{s^{2}}{4R^{2}}}}\right) \right] ,
\end{equation}
where \( \left| hRs\right| <1 \). \( T_{n}(s) \) is then given by \begin{equation}
T_{n}(s)=\frac{1}{n!}\left( \frac{\partial }{\partial h}\right) ^{n}F_{2}(h=0,s).
\end{equation}
 It follows from the previous results that \( P_{n}(s) \) can be expressed
in terms of two unique elementary functions, \( F_{1}(h,s) \) in Eq.~(\ref{eq_generating_F1})
and \( F_{2}(h,s) \) in Eq.~(\ref{eq_generating_F2}), such that\begin{equation}
P_{n}(s)=\frac{\frac{1}{n!}\left( \frac{\partial }{\partial h}\right) ^{n}F_{2}(h=0,s)}{\frac{1}{2n}B(\frac{n}{2}+\frac{1}{2},\frac{1}{2})R^{2n}}=\frac{s^{n-1}\frac{1}{n!}\left( \frac{\partial }{\partial h}\right) ^{n}F_{1}(h=0,s)}{\frac{1}{2n}B(\frac{n}{2}+\frac{1}{2},\frac{1}{2})R^{2n}}.
\end{equation}

It is convenient to summarize the different expressions we have obtained for
the PDF of a uniform \( n \)-dimensional sphere with radius \( R \): 

\begin{enumerate}
\item \textsl{Integral representation}:\begin{equation}
P_{n}(s)=\frac{s^{n-1}\int _{\frac{s}{2}}^{R}\left( R^{2}-x^{2}\right) ^{\frac{n-1}{2}}dx}{\frac{1}{2n}B(\frac{n}{2}+\frac{1}{2},\frac{1}{2})R^{2n}},
\end{equation}
 
\item \textsl{Normalized incomplete beta function representation}:\begin{equation}
P_{n}(s)=n\frac{s^{n-1}}{R^{n}}I_{1-\frac{s^{2}}{4R^{2}}}\left( \frac{n}{2}+\frac{1}{2},\frac{1}{2}\right) ,
\end{equation}

\item \textsl{Incomplete beta function representation}:\begin{equation}
P_{n}(s)=n\frac{s^{n-1}B_{1-\frac{s^{2}}{4R^{2}}}\left( \frac{n}{2}+\frac{1}{2},\frac{1}{2}\right) }{B\left( \frac{n}{2}+\frac{1}{2},\frac{1}{2}\right) R^{n}},
\end{equation}

\item \textsl{Odd-integer finite series expansion representation (\( n= \) odd)}:\begin{equation}
P_{n}(s)=n\times \frac{s^{n-1}}{R^{n}}\frac{n!!}{(n-1)!!}\sum _{i=0}^{\frac{n-1}{2}}\frac{(-1)^{i}}{2i+1}\frac{\left( \frac{n-1}{2}\right) !}{i!\left( \frac{n-1}{2}-i\right) !}\left[ 1-\left( \frac{s}{2R}\right) ^{2i+1}\right] ,
\end{equation}

\item \textsl{Even-integer finite series expansion representation (\( n= \) even)}:\begin{equation}
P_{n}(s)=n\times \frac{s^{n-1}}{R^{n}}\left[ \frac{2}{\pi }\cos ^{-1}\left( \frac{s}{2R}\right) -\frac{s}{\pi }\sum _{i=1}^{\frac{n}{2}}\frac{(n-2i)!!}{(n-2i+1)!!}\left( R^{2}-\frac{s^{2}}{4}\right) ^{\frac{n-2i+1}{2}}R^{2i-2-n}\right] ,
\end{equation}

\item \textsl{Infinite series expansion representation}:\begin{equation}
P_{n}(s)=\frac{s^{n-1}\sum _{i=0}^{\infty }\left( -1\right) ^{i}\frac{2^{i}}{2i+1}\frac{\left( 2i-1\right) !!}{\left( 2i\right) !!}\frac{\left( \frac{n-1}{2}\right) !}{\left( \frac{n-1}{2}-i\right) !}\left[ 1-\left( \frac{s}{2R}\right) ^{2i+1}\right] }{\frac{1}{2n}B\left( \frac{n}{2}+\frac{1}{2},\frac{1}{2}\right) R^{n}},
\end{equation}

\item \textsl{Generating function representation I}:\begin{eqnarray}
P_{n}(s) & = & \frac{\frac{1}{n!}\left( \frac{\partial }{\partial h}\right) ^{n}_{h=0}\left\{ \left( \frac{1}{s}\right) \frac{1}{\sqrt{1-h^{2}R^{2}s^{2}}}\left[ \sin ^{-1}(hRs)-\sin ^{-1}\left( \frac{hRs-\sqrt{1-\frac{s^{2}}{4R^{2}}}}{1-hRs\sqrt{1-\frac{s^{2}}{4R^{2}}}}\right) \right] \right\} }{\frac{1}{2n}B(\frac{n}{2}+\frac{1}{2},\frac{1}{2})R^{2n}},
\end{eqnarray}
 
\item \textsl{Generating function representation II}:\begin{equation}
P_{n}(s)=\frac{s^{n-1}\frac{1}{n!}\left( \frac{\partial }{\partial h}\right) ^{n}_{h=0}\left\{ \frac{1}{\sqrt{1-h^{2}R^{2}}}\left[ \sin ^{-1}(hR)-\sin ^{-1}\left( \frac{hR-\sqrt{1-\frac{s^{2}}{4R^{2}}}}{1-hR\sqrt{1-\frac{s^{2}}{4R^{2}}}}\right) \right] \right\} }{\frac{1}{2n}B(\frac{n}{2}+\frac{1}{2},\frac{1}{2})R^{2n}},
\end{equation}

\item \textsl{Hypergeometric function representation}:\begin{equation}
P_{n}(s)=\frac{2n}{B\left( \frac{n}{2}+\frac{1}{2},\frac{1}{2}\right) }\frac{s^{n-1}}{R^{n+1}}\left[ R\, \, _{2}F_{1}\left( \frac{1}{2},\frac{1}{2}-\frac{n}{2};\frac{3}{2};1\right) -\frac{s}{2}\, \, _{2}F_{1}\left( \frac{1}{2},\frac{1}{2}-\frac{n}{2};\frac{3}{2};\frac{s^{2}}{4R^{2}}\right) \right] ,
\end{equation}
 where \begin{equation}
\int \left( R^{2}-x^{2}\right) ^{\frac{n-1}{2}}dx=R^{n-1}x\, \, _{2}F_{1}\left( \frac{1}{2},\frac{1}{2}-\frac{n}{2};\frac{3}{2};\frac{x^{2}}{R^{2}}\right) ,
\end{equation}
 and \begin{eqnarray}
y(x) & \equiv  & \, \, _{2}F_{1}(a,b;c;x)\nonumber \\
 & = & 1+\frac{ab}{c}\frac{x}{1!}+\frac{a(a+1)b(b+1)}{c(c+1)}\frac{x^{2}}{2!}+\frac{a(a+1)(a+2)b(b+1)(b+2)}{c(c+1)(c+2)}\frac{x^{3}}{3!}+\cdots ,
\end{eqnarray}
 is one of the solutions for the hypergeometric equation\begin{equation}
x(1-x)y''(x)+\left[ c-(a+b+1)x\right] y'(x)-aby(x)=0.
\end{equation}

\end{enumerate}
Using the previous results one can obtain a number of identities and recursion
relations for the probability density functions \( P_{n}(s) \), as we discuss
in Appendix~\ref{appendix_002}.

\section{Spherically symmetric density distributions}
\label{sec_spherical_density}

In this section we generalize the previous results to the case of an \( n \)-dimensional
sphere of radius \( R \) with a variable (but spherically symmetric) density
distribution of the form \( \rho =\rho (r) \), where \( r=\sqrt{x_{1}^{2}+x_{2}^{2}+\cdots x_{n}^{2}} \)
is measured from the center and \[
x_{1}^{2}+x_{2}^{2}+\cdots +x_{n}^{2}\leq R^{2}.\]

As before we begin with the example of a circle (\( n=2 \)) and generalize
to a sphere (\( n\geq 3 \)) later. Following the derivation presented in the
previous section, the positive \( \hat{x} \) direction is chosen to specify
the distribution of those \( \vec{s} \) vectors that are aligned along the
positive \( \hat{x} \) direction. At this stage we must consider the differences
between uniform and non-uniform density distributions. For a given \( s, \)
if point \( 2 \) carries the density information \( \rho (x,y) \), then point
\( 1 \) should have the density information \( \rho (x-s,y). \) It follows
that to incorporate the effects of a spherically symmetric density distribution
the following substitution should be made: \begin{equation}
\label{eq_density_2d}
\rho (\vec{r}_{2})\times \rho (\vec{r}_{1})\rightarrow \rho (x,y)\times \rho (x-s,y).
\end{equation}
 Since the density distributions considered are spherically symmetric, the probability
of finding a given \( s \) in any orientation is still proportional to \( 2\pi s. \)
The PDF \( P_{2}(s) \) can then be expressed in the form 

\begin{equation}
P_{2}(s)=\frac{n_{d}(s)\times \left( n_{1}(s)+n_{2}(s)\right) }{\int _{0}^{2R}n_{d}(s)\times \left( n_{1}(s)+n_{2}(s)\right) ds},
\end{equation}
 where \begin{eqnarray}
n_{d}(s) & = & 2\pi s,\nonumber \\
n_{1}(s) & = & \int _{s-R}^{\frac{s}{2}}dx\int _{-\sqrt{R^{2}-(x-s)^{2}}}^{\sqrt{R^{2}-(x-s)^{2}}}\rho (x,y)\times \rho (x-s,y)dy,\nonumber \\
n_{2}(s) & = & \int _{\frac{s}{2}}^{R}dx\int _{-\sqrt{R^{2}-x^{2}}}^{\sqrt{R^{2}-x^{2}}}\rho (x,y)\times \rho (x-s,y)dy.
\end{eqnarray}
 Substituting \( x-s=x' \) and using \( \rho (x,y)=\rho (-x,y) \) it can be
shown that \( n_{1}(s)=n_{2}(s). \) The expression for \( P_{2}(s) \) can
then be simplified to read 

\begin{eqnarray}
P_{2}(s) & = & \frac{s\int _{\frac{s}{2}}^{R}dx\int _{-\sqrt{R^{2}-x^{2}}}^{\sqrt{R^{2}-x^{2}}}\rho (x,y)\times \rho (x-s,y)dy}{\int _{0}^{2R}\left( s\int _{\frac{s}{2}}^{R}dx\int _{-\sqrt{R^{2}-x^{2}}}^{\sqrt{R^{2}-x^{2}}}\rho (x,y)\times \rho (x-s,y)dy\right) ds}.\label{eq_non_uniform_2} 
\end{eqnarray}

The formalism leading to Eq.~(\ref{eq_non_uniform_2}) can be extended to a
\( 3 \)-dimensional sphere of radius \( R \). For a given \( s \), the \( z \)-axis
is chosen arbitrarily as our reference axis to examine the distribution of those
\( \vec{s} \) vectors that are aligned along the positive \( \hat{z} \) direction.
If the density at point \( 2 \) is \( \rho (x,y,z), \) then the density at
point \( 1 \) will be \( \rho (x,y,z-s). \) In analogy with the \( 2 \)-dimensional
case discussed above, the expression in Eq.~(\ref{eq_density_2d}) must be replaced
by \begin{equation}
\label{eq_density_3d}
\rho (\vec{r}_{2})\times \rho (\vec{r}_{1})=\rho (x,y,z)\times \rho (x,y,z-s).
\end{equation}
 Since the density distributions considered here are spherically symmetric,
the probability of finding a given \( s \) in any orientation is proportional
to \( 4\pi s^{2} \). Hence \( P_{3}(s) \) can be expressed as \begin{eqnarray}
P_{3}(s) & = & \frac{s^{2}\int _{\frac{s}{2}}^{R}dz\int _{-\sqrt{R^{2}-z^{2}}}^{\sqrt{R^{2}-z^{2}}}dx\int _{-\sqrt{R^{2}-z^{2}-x^{2}}}^{\sqrt{R^{2}-z^{2}-x^{2}}}\rho (x,y,z)\times \rho (x,y,z-s)dy}{\int _{0}^{2R}\left( s^{2}\int _{\frac{s}{2}}^{R}dz\int _{-\sqrt{R^{2}-z^{2}}}^{\sqrt{R^{2}-z^{2}}}dx\int _{-\sqrt{R^{2}-z^{2}-x^{2}}}^{\sqrt{R^{2}-z^{2}-x^{2}}}\rho (x,y,z)\times \rho (x,y,z-s)dy\right) ds}\\
 & = & \frac{s^{2}\int ^{R}_{\frac{s}{2}}dx_{3}\int _{-\sqrt{R^{2}-x_{3}^{2}}}^{\sqrt{R^{2}-x_{3}^{2}}}dx_{1}\int _{-\sqrt{R^{2}-x_{3}^{2}-x_{1}^{2}}}^{\sqrt{R^{2}-x_{3}^{2}-x_{1}^{2}}}dx_{2}\rho \left( x_{1},x_{2},x_{3}\right) \rho \left( x_{1},x_{2},x_{3}'\right) }{\int _{0}^{2R}s^{2}ds\int ^{R}_{\frac{s}{2}}dx_{3}\int _{-\sqrt{R^{2}-x_{3}^{2}}}^{\sqrt{R^{2}-x_{3}^{2}}}dx_{1}\int _{-\sqrt{R^{2}-x_{3}^{2}-x_{1}^{2}}}^{\sqrt{R^{2}-x_{3}^{2}-x_{1}^{2}}}dx_{2}\rho \left( x_{1},x_{2},x_{3}\right) \rho \left( x_{1},x_{2},x_{3}'\right) },\label{eq_pdf_3d_001} 
\end{eqnarray}
 where \( x_{3}'=x_{3}-s \).

\begin{figure}
\phantom{aaa}
\vfill
{\par\centering \resizebox*{6in}{!}{\includegraphics{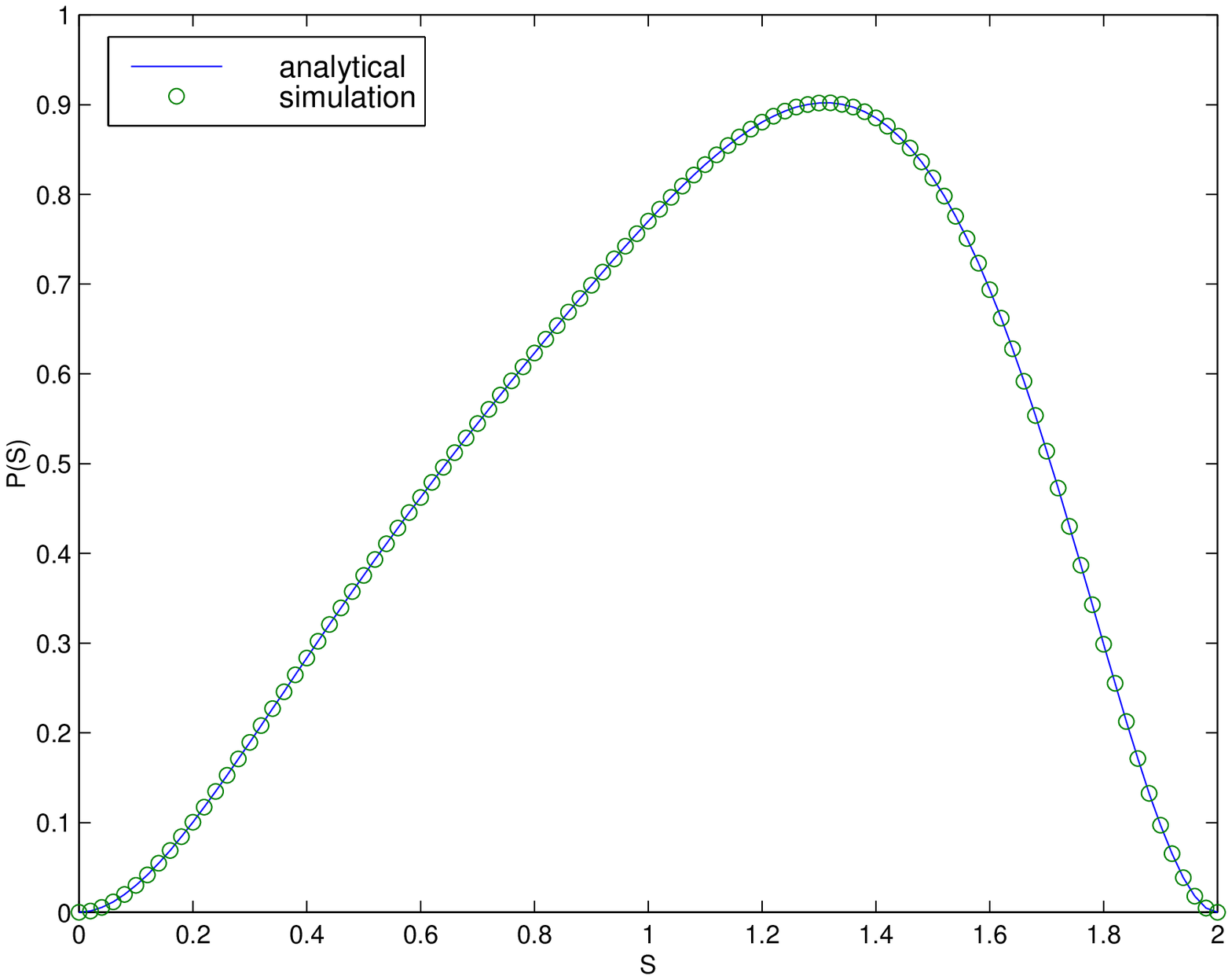}} \par}
\caption{PDF for a \protect\( 3\protect \)-dimensional sphere of radius \protect\( R=1\protect \)
and \protect\( \rho \propto r^{2}\protect \).\label{fig_sys_aaa}}
\vfill\phantom{aaa}\end{figure}

\begin{figure}
\phantom{aaa}
\vfill
{\par\centering \resizebox*{6in}{!}{\includegraphics{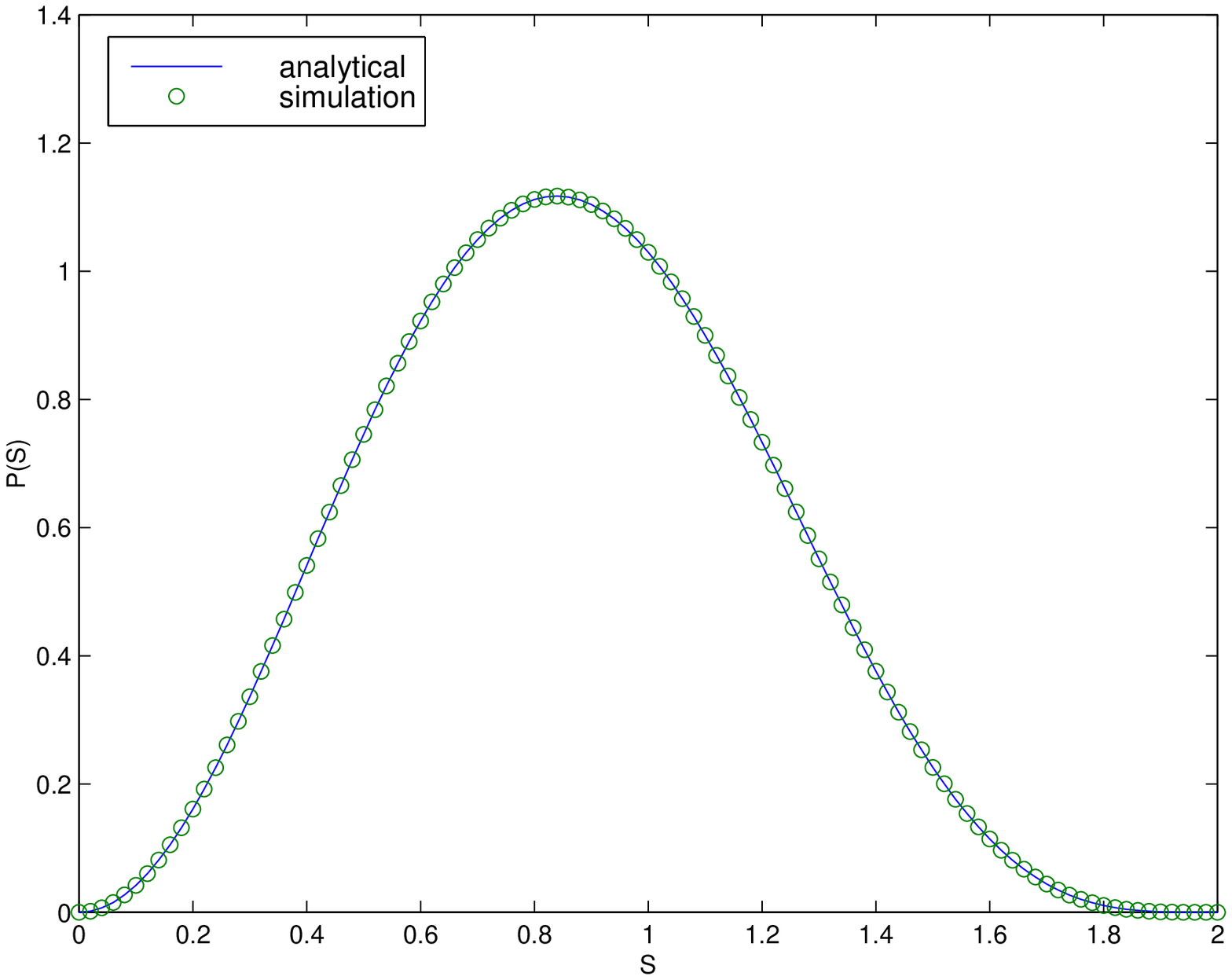}} \par}
\caption{PDF for a \protect\( 3\protect \)-dimensional sphere of radius \protect\( R=1\protect \)
and \protect\( \rho \propto 1-r^{2}\protect \).\label{fig_sys_bbb}}
\vfill\phantom{aaa}\end{figure}

Up to this point our discussion has been completely general. To continue we
next evaluate \( P_{3}(s) \) for a \( 3 \)-dimensional sphere of radius \( R \)
using two different spherically symmetric density distributions. Consider first
\begin{equation}
\label{eq_example_1}
\rho (r)=\frac{5N}{4\pi R^{5}}r^{2},
\end{equation}
 where \( N=4\pi \int _{0}^{R}r^{2}\rho (r)dr \), and \( r \) is measured
from the center of the spherical distribution. Combining Eqs.~(\ref{eq_pdf_3d_001})
and (\ref{eq_example_1}) we find \begin{equation}
\label{eq_sys_aaa}
P_{3}(s)=\frac{25}{7}\frac{s^{2}}{R^{3}}-\frac{25}{4}\frac{s^{3}}{R^{4}}+5\frac{s^{4}}{R^{5}}-\frac{25}{16}\frac{s^{5}}{R^{6}}+\frac{5}{448}\frac{s^{9}}{R^{10}}.
\end{equation}
 A plot of Eq.~(\ref{eq_sys_aaa}) when \( R=1 \), along with the corresponding
Monte Carlo results, is shown in Fig.~\ref{fig_sys_aaa}. The second spherically
symmetric distribution we consider is \begin{equation}
\label{eq_example_2}
\rho (r)=\frac{N}{4\pi \left( \frac{1}{3}-\frac{\alpha }{5}\right) R^{3}}\left[ 1-\alpha \left( \frac{r}{R}\right) ^{2}\right] ,
\end{equation}
 where \( N=4\pi \int _{0}^{R}r^{2}\rho (r)dr \), \( 0\leq \alpha \leq 1 \),
and \( r \) is measured from the center. Combining Eqs.~(\ref{eq_pdf_3d_001})
and (\ref{eq_example_2}) we find\begin{eqnarray}
P_{3}(s) & = & \frac{15(35-42\alpha +15\alpha ^{2})s^{2}}{7(5-3\alpha )^{2}R^{3}}-\frac{225(1-\alpha )^{2}s^{3}}{4(5-3\alpha )^{2}R^{4}}-\frac{15\alpha s^{4}}{(5-3\alpha )R^{5}}\nonumber \\
 &  & +\frac{75(1+6\alpha -3\alpha ^{2})s^{5}}{16(5-3\alpha )^{2}R^{6}}-\frac{15\alpha s^{7}}{8(5-3\alpha )^{2}R^{8}}+\frac{45\alpha ^{2}s^{9}}{448(5-3\alpha )^{2}R^{10}}.\label{eq_sys_bbb} 
\end{eqnarray}
 A plot of Eq.~(\ref{eq_sys_bbb}) when \( R=\alpha =1 \) is shown in Fig.~\ref{fig_sys_bbb},
along with the corresponding Monte Carlo results.

A general formula for the probability density function for an \( n \)-dimensional
sphere with radius \( R \) having a spherically symmetric density distribution
can be derived from the previous results. We find \begin{equation}
\label{eq_spherical_001}
P_{n}(s)=\frac{s^{n-1}\int _{\frac{s}{2}}^{R}dx_{n}\int _{-\sqrt{R^{2}-x_{n}^{2}}}^{\sqrt{R^{2}-x_{n}^{2}}}dx_{1}\cdots \cdots \int _{-\sqrt{R^{2}-x_{n}^{2}-x_{1}^{2}-\cdots x_{n-1}^{2}}}^{\sqrt{R^{2}-x_{n}^{2}-x_{1}^{2}-\cdots x_{n-1}^{2}}}\rho \left( \mathbf{X}\right) \rho \left( \mathbf{X}'\right) dx_{n-1}}{\int _{0}^{2R}\left[ s^{n-1}\int _{\frac{s}{2}}^{R}dx_{n}\int _{-\sqrt{R^{2}-x_{n}^{2}}}^{\sqrt{R^{2}-x_{n}^{2}}}dx_{1}\cdots \cdots \int _{-\sqrt{R^{2}-x_{n}^{2}-x_{1}^{2}-\cdots x_{n-1}^{2}}}^{\sqrt{R^{2}-x_{n}^{2}-x_{1}^{2}-\cdots x_{n-1}^{2}}}\rho \left( \mathbf{X}\right) \rho \left( \mathbf{X}'\right) dx_{n-1}\right] ds},
\end{equation}
 where\begin{eqnarray}
\rho (\mathbf{X}) & = & \rho (x_{1},x_{2},x_{3},\cdots \cdots ,x_{n}),\nonumber \\
\rho (\mathbf{X}') & = & \rho (x_{1},x_{2},x_{3},\cdots \cdots ,x_{n}-s).
\end{eqnarray}

Another density distribution we wish to study is a Gaussian. As an example,
consider the case of an \( n \)-dimensional sphere of radius \( R\rightarrow \infty  \)
with a Gaussian density distribution \( \rho (r) \) given by \begin{equation}
\label{eq_n_gaussian_density}
\rho _{n}(r)=\frac{N}{(2\pi )^{\frac{n}{2}}\sigma ^{n}}e^{-\frac{1}{2}\frac{r^{2}}{\sigma ^{2}}},
\end{equation}
 where\begin{equation}
\label{eq_normalization_gaussian}
N=\lim _{R\rightarrow \infty }n\frac{\pi ^{\frac{n}{2}}}{\Gamma \left( \frac{n}{2}+1\right) }\int _{0}^{R}\rho _{n}(r)r^{n-1}dr.
\end{equation}
 In Eq.~(\ref{eq_normalization_gaussian}) \( r \) is measured from the center
of the spherical distribution and the integral is over all space. Recall that
\begin{eqnarray}
\int _{0}^{\infty }x^{n}e^{-\frac{x^{2}}{2\sigma ^{2}}}dx & = & 2^{\frac{n-1}{2}}\Gamma \left( \frac{n+1}{2}\right) \sigma ^{n+1}.\label{eq_identity_001} 
\end{eqnarray}
 Combining Eqs.~(\ref{eq_n_gaussian_density}) and (\ref{eq_identity_001}),
the PDF for an \( n \)-dimensional sphere in an infinite space with a Gaussian
density distribution can be expressed as\begin{equation}
\label{eq_n_pdf_gaussian}
P_{n}(s)=\lim _{R\rightarrow \infty }\frac{s^{n-1}e^{-\frac{s^{2}}{4\sigma ^{2}}}}{\int _{0}^{2R}s^{n-1}e^{-\frac{s^{2}}{4\sigma ^{2}}}ds}=\frac{1}{2^{n-1}\Gamma \left( \frac{n}{2}\right) \sigma ^{n}}s^{n-1}e^{-\frac{s^{2}}{4\sigma ^{2}}}.
\end{equation}
 For \( n=3 \), Eq.~(\ref{eq_n_pdf_gaussian}) agrees with the result obtained
earlier in Ref.~\cite{Schleef}. Finally, we note that the maximum probability,
denoted by \( S_{max} \), occurs at \begin{equation}
S_{max}=\sqrt{2(n-1)}\, \sigma .
\end{equation}

\section{Arbitrary density distributions}
\label{sec_arbitrary_density}

We consider in this section the probability density functions for an \( n \)-dimensional
sphere of radius \( R \) having an arbitrary density distribution,

\begin{equation}
\label{eqn_4_001}
\rho =\rho (\mathbf{X})=\rho (x_{1},x_{2},\cdots ,x_{n}),
\end{equation}
 where \begin{equation}
x^{2}_{1}+x^{2}_{2}+\cdots +x^{2}_{n}\leq R^{2},
\end{equation}
 The proportionality factors, \( 2\pi s \) (\( n=2 \))  and \( 4\pi s^{2} \)
(\( n=3 \)), cannot be applied here directly because the density function is
not spherically symmetric. For a given \( s, \) each direction of \( \vec{s} \)
carries different information specified by the density distribution function. 

\begin{figure}
\phantom{aaa}
\vfill
{\par\centering \includegraphics{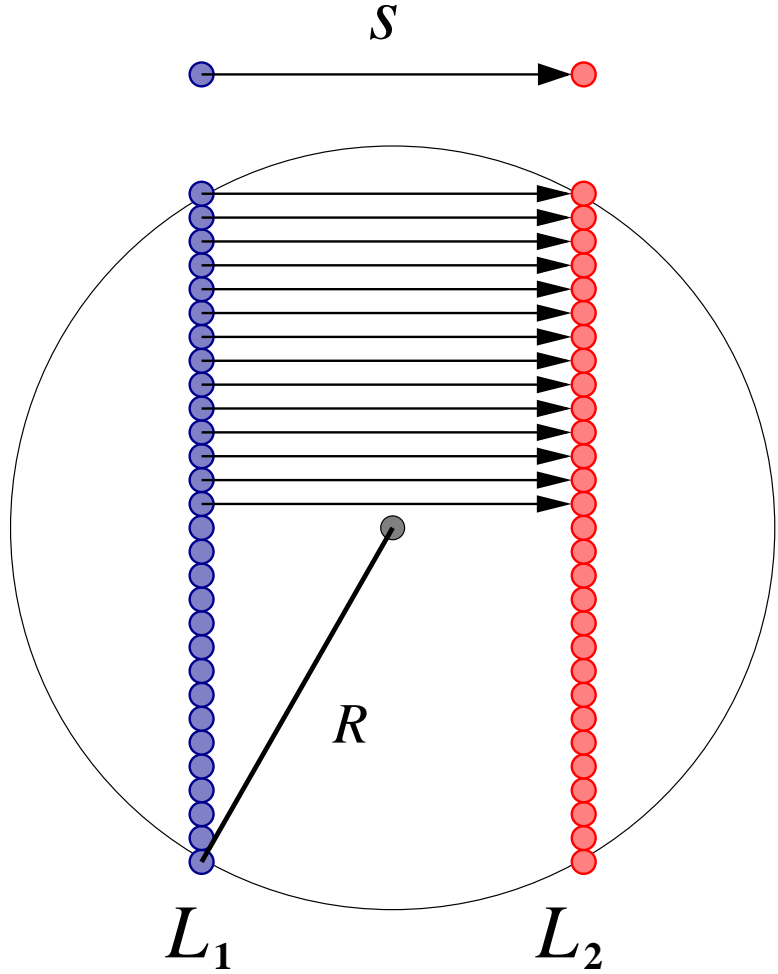} \par}
\caption{Locus of points in separated by a fixed distance \protect\( s=|s|\, \hat{x}\protect \)
from a line \protect\( L_{1}\protect \) in a circle of radius \protect\( R\protect \).
If \protect\( L_{1}\protect \) is the line \protect\( x=-s/2\protect \) then
\protect\( L_{2}\protect \) is the line \protect\( x=+s/2\protect \), where
\protect\( -\left( R^{2}-s^{2}/4\right) ^{1/2}\leq y\leq \left( R^{2}-s^{2}/4\right) ^{1/2}\protect \).
\label{fig_choice_1}}
\vfill\phantom{aaa}\end{figure}

\begin{figure}
\phantom{aaa}
\vfill
{\par\centering \resizebox*{6.25in}{!}{\includegraphics{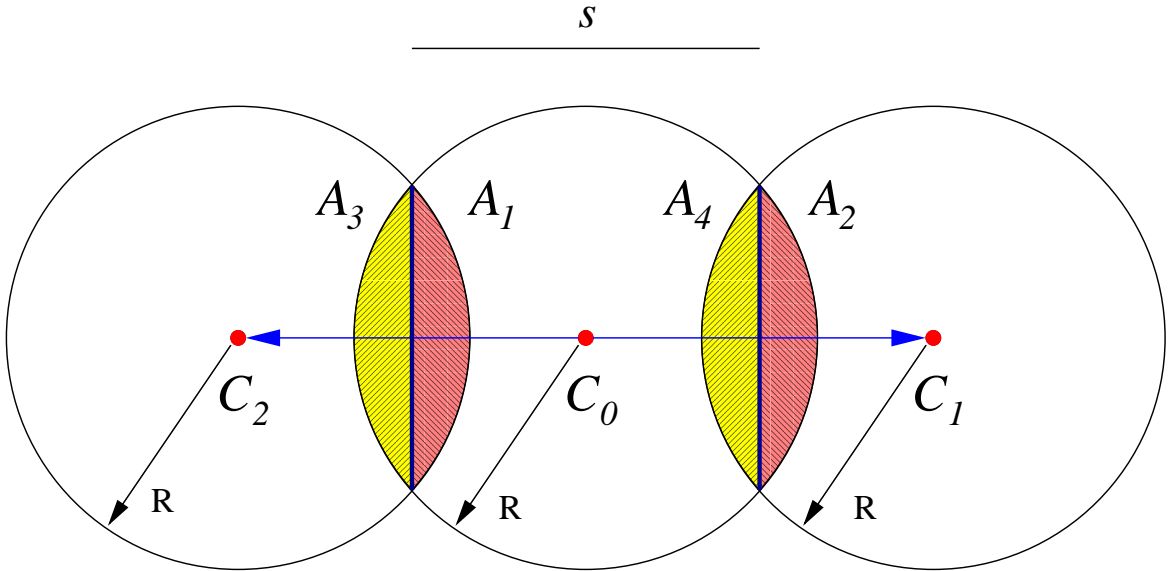}} \par}
\caption{Geometric interpretation of the probability density function \protect\( P_{2}(s)\protect \).
From Figs.~\ref{fig_choice_1} and \ref{fig_choice_2}, the probability that
two points are separated by a distance \protect\( s\protect \) (\protect\( \left| \vec{s}\, \right| =\left| s\hat{x}\right| \protect \))
in a circle of radius \protect\( R\protect \) is proportional to the total
shaded area, \protect\( A_{3}\bigcup A_{1}\bigcup A_{4}\bigcup A_{2}\protect \)
, in the circle \protect\( C_{0}\protect \). As can be seen from this figure,
this shaded area is given by the overlap of \protect\( C_{0}\protect \) with
two identical circles \protect\( C_{1}\protect \) and \protect\( C_{2}\protect \)
as shown, where \protect\( C_{0}\protect \) is the circle \protect\( x^{2}+y^{2}=R^{2}\protect \),
\protect\( C_{1}\protect \) is \protect\( (x-s)^{2}+y^{2}=R^{2}\protect \),
and \protect\( C_{2}\protect \) is \protect\( (x+s)^{2}+y^{2}=R^{2}\protect \).
\label{fig_shift_center}}
\vfill\phantom{aaa}\end{figure}

We begin with a circle of radius \( R \) and the conventional notation for
polar coordinates, \( x=r\cos \phi  \) and \( y=r\sin \phi  \). For a given
\( \vec{s}=\vec{r}_{2}-\vec{r}_{1}, \) the PDF, \( P_{2}(s), \) is proportional
to \( A_{\vec{s}}\times \rho (\vec{r}_{2})\times \rho (\vec{r}_{1}), \) where
\( A_{\vec{s}} \) is the overlapping area between the original circle and a
second identical one whose center is shifted to \( \vec{s} \), as described
in the previous sections. In \( 2 \)-dimensional space \( \vec{s} \) can be
characterized by an angle \( \phi  \) in the range \( 0\leq \phi \leq 2\pi  \).
One can understand the new features that arise for a non-uniform density distribution
by referring back to Fig.~\ref{fig_choice_1}. In the case of a uniform density
distribution the picture formed by the vector \( \vec{s} \) extending between
\( L_{1} \) and \( L_{2} \) is unchanged by a rotation of the entire pattern
about the positive \( x \)-axis. However, for a non-uniform distribution the
effect of such a rotation is to shift the vectors into a new region for which
the density of points is not the same as it was initially. Stated another way,
for a fixed \( \left| \vec{s}\, \right|  \) the shape of the overlapping area
or volume is the same, but will contain a different fraction of the points depending
on the orientation of \( \vec{s} \). To deal with this effect, one can rotate
the coordinate system so that the pattern remains as shown in Fig.~\ref{fig_choice_1},
but with an appropriately transformed density distribution. To specify this
transformation, we associate \( \vec{s} \) with a rotation operator \( \mathbf{R}(\vec{s}\, ) \)
such that the direction of \( \vec{s} \) is the new \( \hat{x} \) direction
where \( \hat{s}=\hat{x}' \), \( \hat{x}'\cdot \hat{x}=\cos \phi  \), \( \hat{x}'\cdot \hat{y}=\sin \phi  \),
\( \hat{y}'\cdot \hat{x}=-\sin \phi  \) and \( \hat{y}'\cdot \hat{y}=\cos \phi  \).
We utilize the transformation matrix for a \( 2 \)-dimensional rotation \begin{equation}
R_{2\times 2}(\phi )=\left[ \begin{array}{cc}
\cos \phi  & \sin \phi \\
-\sin \phi  & \cos \phi 
\end{array}\right] 
\end{equation}
 to describe \( \mathbf{R}(\vec{s}\, ) \). Notice that \( R_{2\times 2}(\phi ) \)
is an orthogonal matrix which satisfies \( R_{2\times 2}^{-1}(\phi )=R_{2\times 2}^{T}(\phi ) \),
where \( T \) denotes the transpose. Recall that the functional form of the
density distribution in Eq.~(\ref{eqn_4_001}) is written in the original coordinate
system. The inverse transformation matrix, \( R_{2\times 2}^{-1}(\phi )=R_{2\times 2}^{T}(\phi ) \),
should be used to transmit the correct density information to the new coordinate
system \( \hat{s} \) (\( \hat{x}' \)).

It is convenient to introduce the following general notations, \begin{eqnarray}
\vec{\mathbf{X}}' & = & R_{2\times 2}^{T}(\phi )\vec{\mathbf{X}},\\
\left[ \begin{array}{c}
x'\\
y'
\end{array}\right]  & = & \left[ \begin{array}{cc}
\cos \phi  & -\sin \phi \\
\sin \phi  & \cos \phi 
\end{array}\right] \left[ \begin{array}{c}
x\\
y
\end{array}\right] ,
\end{eqnarray}
 and \begin{eqnarray}
\vec{\mathbf{X}}'' & = & R_{2\times 2}^{T}(\phi )(\vec{\mathbf{X}}-\vec{\mathbf{S}}),\\
\left[ \begin{array}{c}
x''\\
y''
\end{array}\right]  & = & \left[ \begin{array}{cc}
\cos \phi  & -\sin \phi \\
\sin \phi  & \cos \phi 
\end{array}\right] \left[ \begin{array}{c}
x-s\\
y
\end{array}\right] ,
\end{eqnarray}
where \begin{equation}
\vec{\mathbf{X}}=\left[ \begin{array}{c}
x\\
y
\end{array}\right] ,\, \, \, \vec{\mathbf{X}}'=\left[ \begin{array}{c}
x'\\
y'
\end{array}\right] ,\, \, \, \vec{\mathbf{X}}''=\left[ \begin{array}{c}
x''\\
y''
\end{array}\right] ,\, \, \, \vec{\mathbf{S}}=\left[ \begin{array}{c}
s\\
0
\end{array}\right] .
\end{equation}
We can then express the PDF for a circle of radius \( R \) with an arbitrary
density distribution as \begin{equation}
P_{2}(s)=\frac{s\int _{0}^{2\pi }d\phi \int _{\frac{s}{2}}^{R}dx\int _{-\sqrt{R^{2}-x^{2}}}^{\sqrt{R^{2}-x^{2}}}\rho (\mathbf{X}')\rho (\mathbf{X}'')dy}{\int _{0}^{2R}\left[ s\int _{0}^{2\pi }d\phi \int _{\frac{s}{2}}^{R}dx\int _{-\sqrt{R^{2}-x^{2}}}^{\sqrt{R^{2}-x^{2}}}\rho (\mathbf{X}')\rho (\mathbf{X}'')dy\right] ds},
\end{equation}
 where \begin{equation}
\rho (\mathbf{X}')=\rho (\cos \phi \, x-\sin \phi \, y,\, \sin \phi \, x+\cos \phi \, y),
\end{equation}
\begin{equation}
\rho (\mathbf{X}'')=\rho (\cos \phi \, (x-s)-\sin \phi \, y,\, \sin \phi \, (x-s)+\cos \phi \, y).
\end{equation}

\begin{figure}
\phantom{aaa}
\vfill
{\par\centering \resizebox*{6in}{!}{\includegraphics{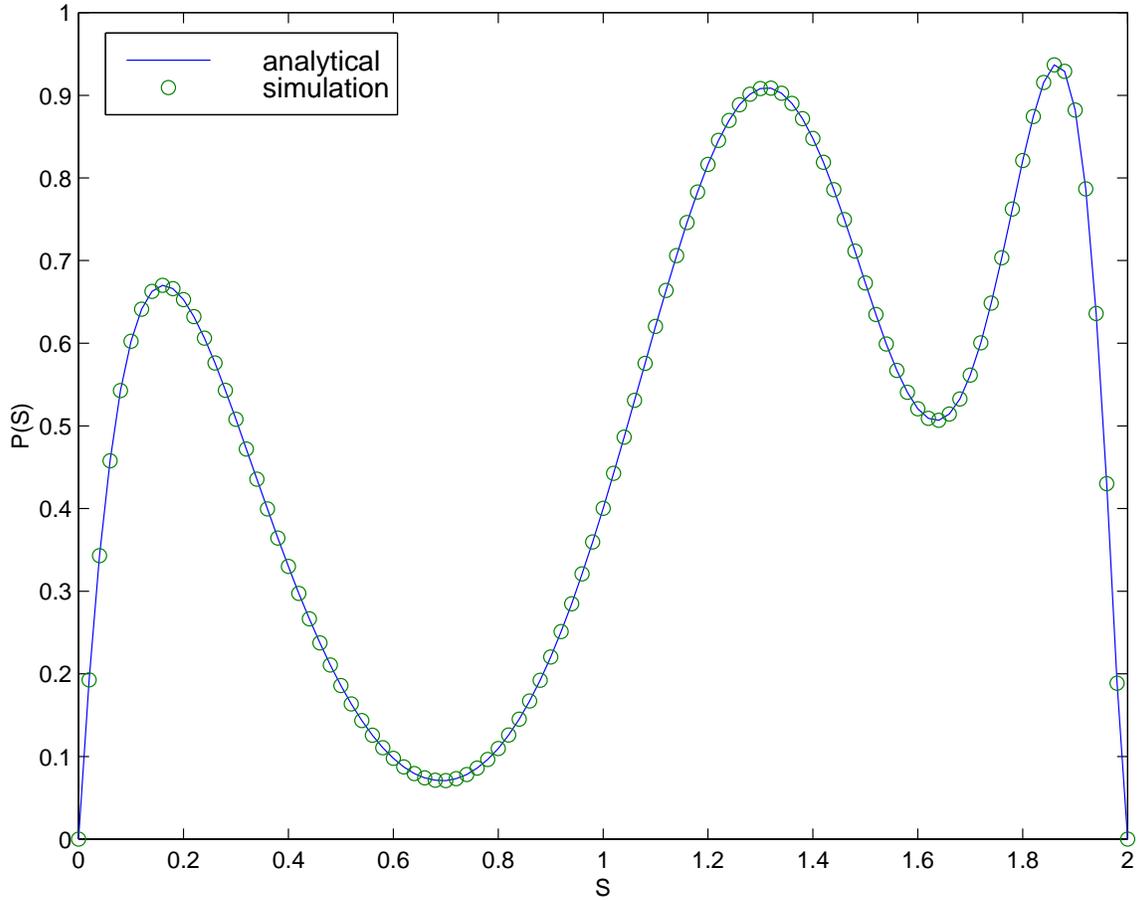}} \par}
\caption{Plot of \protect\( P_{2}(s)\protect \) as a function of \protect\( s\protect \)
for a \protect\( 2\protect \)-dimensional circle of radius \protect\( R=1\protect \)
for the case \protect\( \rho \propto x^{4}y^{4}\protect \). See text for further
discussion.\label{density_2}}
\vfill\phantom{aaa}\end{figure}

As an example , consider a circle of radius \( R \) and non-uniform density
distribution \( \rho (x,y) \) given by\begin{equation}
\label{eq_example_density_2d}
\rho (x,y)=\frac{640N}{3\pi R^{10}}x^{4}y^{4}=\frac{640N}{3\pi R^{10}}r^{8}\cos ^{4}\phi \sin ^{4}\phi ,
\end{equation}
 where \( N \) is a normalization constant \begin{equation}
N=\int _{-R}^{R}dx\int _{-\sqrt{R^{2}-x^{2}}}^{\sqrt{R^{2}-x^{2}}}\rho (x,y)dy.
\end{equation}
For this example the \( 2 \)-dimensional PDF \( P_{2}(s) \) is then given
by \begin{eqnarray}
P_{2}(s) & = & \frac{875}{81}\frac{s}{R^{2}}+\frac{500}{3}\frac{s^{3}}{R^{4}}+\frac{7400}{21}\frac{s^{5}}{R^{6}}+\frac{400}{3}\frac{s^{7}}{R^{8}}+10\frac{s^{9}}{R^{10}}\nonumber \\
 &  & -\frac{\sqrt{4R^{2}-s^{2}}}{\pi }\left[ f_{1}(s)+f_{2}(s)-f_{3}(s)\right] \nonumber \\
 &  & -\frac{\sin ^{-1}\left( \frac{s}{2R}\right) }{\pi }\left( \frac{1750}{81}\frac{s}{R^{2}}+\frac{1000}{3}\frac{s^{3}}{R^{4}}+\frac{14800}{21}\frac{s^{5}}{R^{6}}+\frac{800}{3}\frac{s^{7}}{R^{8}}+20\frac{s^{9}}{R^{10}}\right) ,
\end{eqnarray}
 where \begin{eqnarray}
f_{1}(s) & = & \frac{14875}{162}\frac{s^{2}}{R^{4}}+\frac{92500}{243}\frac{s^{4}}{R^{6}}+\frac{553985}{1701}\frac{s^{6}}{R^{8}},\\
f_{2}(s) & = & \frac{260315}{10206}\frac{s^{10}}{R^{12}}+\frac{113693}{47628}\frac{s^{14}}{R^{16}}+\frac{2509}{142884}\frac{s^{18}}{R^{20}},\\
f_{3}(s) & = & \frac{2725}{1134}\frac{s^{8}}{R^{10}}+\frac{1438825}{142884}\frac{s^{12}}{R^{14}}+\frac{89189}{285768}\frac{s^{16}}{R^{18}}.
\end{eqnarray}
 Figure~\ref{density_2} exhibits \( P_{2}(s) \) when \( R=1 \), and illustrates
the agreement between the Monte Carlo simulation and the analytical result given
above.

The preceding discussion can be extended to a \( 3 \)-dimensional sphere of
radius \( R \) with an arbitrary density distribution \( \rho =\rho (x,y,z) \),
where \( x=r\sin \theta \cos \phi  \), \( y=r\sin \theta \sin \phi  \), and
\( z=r\cos \theta  \) are the usual \( 3 \)-dimensional spherical coordinates,
and \( x^{2}+y^{2}+z^{2}\leq R^{2} \). For a given \( \vec{s}=\vec{r}_{2}-\vec{r}_{1}, \)
the PDF \( P_{3}(s) \) is proportional to \( \rho (\vec{r}_{2})\times \rho (\vec{r}_{1})\times V_{\vec{s}}, \)
where \( V_{\vec{s}} \) is the overlapping volume between the original sphere
and a second identical one whose center is shifted to \( \vec{s} \). In \( 3 \)-dimensional
space \( \vec{s} \) can be oriented at any angle \( \phi  \) between \( 0 \)
and \( 2\pi  \), and the angle \( \theta  \) can lie between \( 0 \) and
\( \pi  \). A rotation matrix \( R_{3\times 3}(\theta ,\phi ) \) is used to
represent the rotation operator \( \mathbf{R}(\vec{s}\, ) \) associated with
a given \( \vec{s} \) such that \begin{equation}
R_{3\times 3}(\theta ,\phi )=R_{3\times 3}(\theta )\times R_{3\times 3}(\phi )=\left[ \begin{array}{ccc}
\cos \theta \cos \phi \,  & \, \cos \theta \sin \phi \,  & \, -\sin \theta \\
-\sin \phi \,  & \, \cos \phi \,  & \, 0\\
\sin \theta \cos \phi \,  & \, \sin \theta \sin \phi \,  & \, \cos \theta 
\end{array}\right] ,
\end{equation}
 where \begin{eqnarray}
R_{3\times 3}(\theta ) & = & \left[ \begin{array}{ccc}
\cos \theta \,  & \, 0\,  & \, -\sin \theta \\
0\,  & \, 1\,  & \, 0\\
\sin \theta \,  & \, 0\,  & \, \cos \theta 
\end{array}\right] ,\\
R_{3\times 3}(\phi ) & = & \left[ \begin{array}{ccc}
\cos \phi \,  & \, \sin \phi \,  & \, 0\\
-\sin \phi \,  & \, \cos \phi \,  & \, 0\\
0\,  & \, 0\,  & \, 1
\end{array}\right] .
\end{eqnarray}
 We observe the following:  

\begin{enumerate}
\item The rotation matrices \( R_{3\times 3}(\theta ,\phi ) \), \( R_{3\times 3}(\theta ) \),
and \( R_{3\times 3}(\phi ) \) are orthogonal so that \begin{equation}
R_{3\times 3}^{-1}(\theta ,\phi )=R_{3\times 3}^{T}(\theta ,\phi )=R^{T}_{3\times 3}(\phi )R_{3\times 3}^{T}(\theta ).
\end{equation}

\item The purpose of \( R_{3\times 3}(\phi ) \) is to transform the coordinate system
from \( \left( x,y,z\right)  \) to a second coordinate system \( \left( x^{1},y^{1},z^{1}\right)  \)
given by \begin{equation}
\left[ \begin{array}{c}
x^{1}\\
y^{1}\\
z^{1}
\end{array}\right] =\left[ \begin{array}{ccc}
\cos \phi \,  & \, \sin \phi \,  & \, 0\\
-\sin \phi \,  & \, \cos \phi \,  & \, 0\\
0\,  & \, 0\,  & \, 1
\end{array}\right] \left[ \begin{array}{c}
x\\
y\\
z
\end{array}\right] .
\end{equation}

\item The purpose of \( R_{3\times 3}(\theta ) \) is to transform the coordinate
system from \( \left( x^{1},y^{1},z^{1}\right)  \) to a third coordinate system
\( \left( x^{2},y^{2},z^{2}\right)  \) where \begin{equation}
\left[ \begin{array}{c}
x^{2}\\
y^{2}\\
z^{2}
\end{array}\right] =\left[ \begin{array}{ccc}
\cos \theta \,  & \, 0\,  & \, -\sin \theta \\
0\,  & \, 1\,  & \, 0\\
\sin \theta \,  & \, 0\,  & \, \cos \theta 
\end{array}\right] \left[ \begin{array}{c}
x^{1}\\
y^{1}\\
z^{1}
\end{array}\right] .
\end{equation}
 
\end{enumerate}
Define the following notations, 

\begin{eqnarray}
\vec{\mathbf{X}}' & = & R_{3\times 3}^{T}(\theta ,\phi )\vec{\mathbf{X}},\nonumber \\
 & = & R_{3\times 3}^{T}(\phi )R_{3\times 3}^{T}(\theta )\vec{\mathbf{X}},
\end{eqnarray}
 \begin{eqnarray}
\vec{\mathbf{X}}'' & = & R_{3\times 3}^{T}(\theta ,\phi )(\vec{\mathbf{X}}-\vec{\mathbf{S}})\nonumber \\
 & = & R_{3\times 3}^{T}(\phi )R_{3\times 3}^{T}(\theta )(\vec{\mathbf{X}}-\vec{\mathbf{S}}),
\end{eqnarray}
where \begin{equation}
\vec{\mathbf{X}}=\left[ \begin{array}{c}
x\\
y\\
z
\end{array}\right] ,\, \, \, \vec{\mathbf{X}}'=\left[ \begin{array}{c}
x'\\
y'\\
z'
\end{array}\right] ,\, \, \, \vec{\mathbf{X}}''=\left[ \begin{array}{c}
x''\\
y''\\
z''
\end{array}\right] ,\, \, \, \vec{\mathbf{S}}=\left[ \begin{array}{c}
0\\
0\\
1
\end{array}\right] .
\end{equation}
 The PDF \( P_{3}(s) \) can then be expressed in the form \begin{equation}
P_{3}(s)=\frac{s^{2}\int _{0}^{\pi }\sin \theta d\theta \int _{0}^{2\pi }d\phi \int _{\frac{s}{2}}^{R}dz\int _{-\sqrt{R^{2}-z^{2}}}^{\sqrt{R^{2}-z^{2}}}dx\int _{-\sqrt{R^{2}-z^{2}-x^{2}}}^{\sqrt{R^{2}-z^{2}-x^{2}}}\rho (\mathbf{X}')\rho (\mathbf{X}'')dy}{\int _{0}^{2R}\left[ s^{2}\int _{0}^{\pi }\sin \theta d\theta \int _{0}^{2\pi }d\phi \int _{\frac{s}{2}}^{R}dz\int _{-\sqrt{R^{2}-z^{2}}}^{\sqrt{R^{2}-z^{2}}}dx\int _{-\sqrt{R^{2}-z^{2}-x^{2}}}^{\sqrt{R^{2}-z^{2}-x^{2}}}\rho (\mathbf{X}')\rho (\mathbf{X}'')dy\right] ds},
\end{equation}
 where 

\begin{eqnarray}
\rho (\mathbf{X}') & = & \rho (x',y',z'),\nonumber \\
x' & = & \cos \theta \cos \phi \, x-\sin \phi \, y+\sin \theta \cos \phi \, z,\nonumber \\
y' & = & \cos \theta \sin \phi \, x+\cos \phi \, y+\sin \theta \sin \phi \, z,\nonumber \\
z' & = & -\sin \theta \, x+\cos \theta \, z,
\end{eqnarray}
 \begin{eqnarray}
\rho (\mathbf{X}'') & = & \rho (x'',y'',z''),\nonumber \\
x'' & = & \cos \theta \cos \phi \, x-\sin \phi \, y+\sin \theta \cos \phi \, (z-s),\nonumber \\
y'' & = & \cos \theta \sin \phi \, x+\cos \phi \, y+\sin \theta \sin \phi \, (z-s),\nonumber \\
z'' & = & -\sin \theta \, x+\cos \theta \, (z-s).
\end{eqnarray}

\begin{figure}
\phantom{aaa}
\vfill
{\par\centering \resizebox*{6in}{!}{\includegraphics{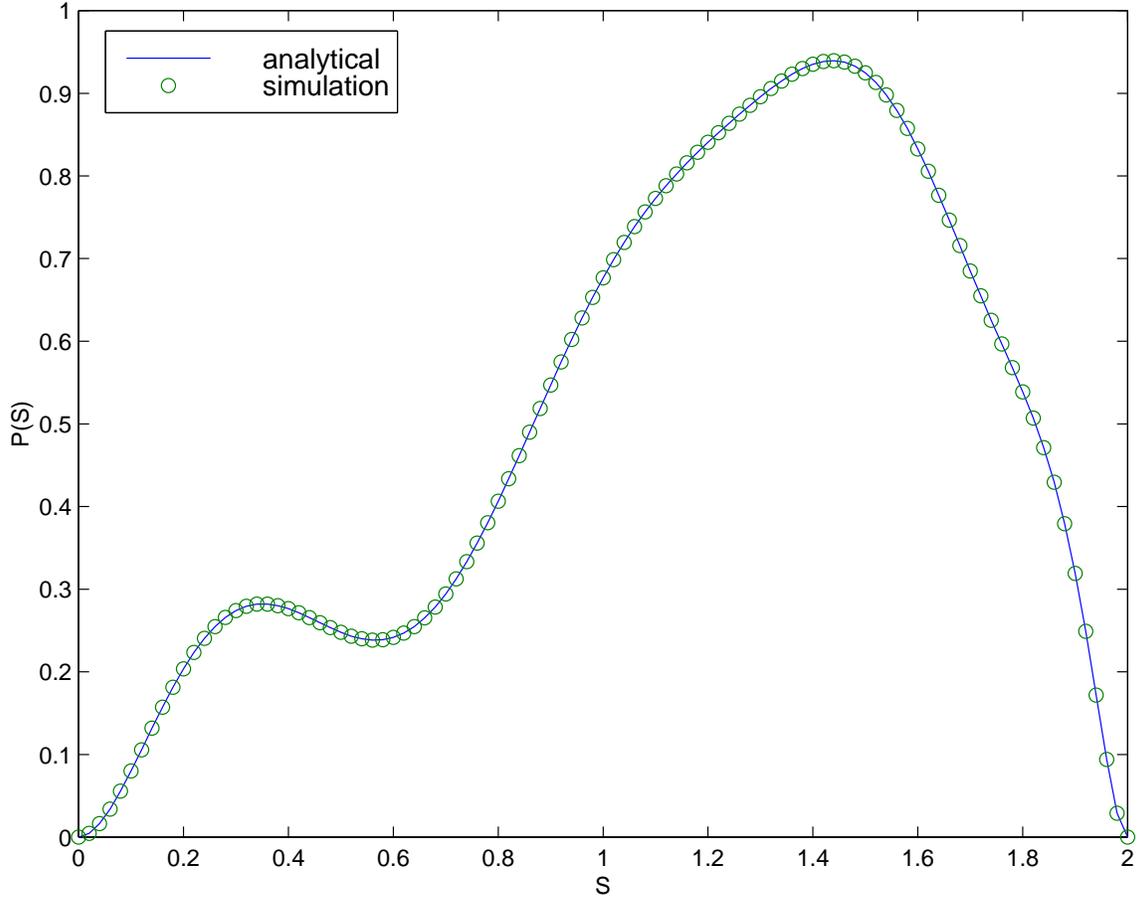}} \par}
\caption{Plot of \protect\( P_{3}(s)\protect \) as a function of \protect\( s\protect \)
for a \protect\( 3\protect \)-dimensional sphere of radius \protect\( R=1\protect \)
and \protect\( \rho (x,y,z)=(945N/4\pi )x^{2}y^{2}z^{2}\protect \).\label{density_3}}
\vfill\phantom{aaa}\end{figure}

As an example, consider a \( 3 \)-dimensional sphere of radius \( R \) and
non-uniform density distribution \( \rho (x,y,z) \) given by \begin{equation}
\label{eq_example_density_3d}
\rho (x,y,z)=\frac{945N}{4\pi R^{9}}x^{2}y^{2}z^{2}=\frac{945N}{4\pi R^{9}}r^{6}\sin ^{4}\theta \cos ^{2}\theta \cos ^{2}\phi \sin ^{2}\phi ,
\end{equation}
 where \( N \) is the normalizing factor \begin{equation}
N=\int _{-R}^{R}dx\int _{-\sqrt{R^{2}-x^{2}}}^{\sqrt{R^{2}-x^{2}}}dy\int _{-\sqrt{R^{2}-x^{2}-y^{2}}}^{\sqrt{R^{2}-x^{2}-y^{2}}}\rho (x,y,z)dz.
\end{equation}
 Then \begin{eqnarray}
P_{3}(s) & = & \frac{1701}{143}\frac{s^{2}}{R^{3}}-\frac{25515}{572}\frac{s^{3}}{R^{4}}+\frac{8505}{143}\frac{s^{4}}{R^{5}}-\frac{8505}{208}\frac{s^{5}}{R^{6}}+\frac{567}{11}\frac{s^{6}}{R^{7}}-\frac{6237}{104}\frac{s^{7}}{R^{8}}+9\frac{s^{8}}{R^{9}}\nonumber \\
 &  & +\frac{201285}{9152}\frac{s^{9}}{R^{10}}-\frac{181629}{18304}\frac{s^{11}}{R^{12}}+\frac{16443}{6656}\frac{s^{13}}{R^{14}}-\frac{6075}{18304}\frac{s^{15}}{R^{16}}+\frac{10899}{585728}\frac{s^{17}}{R^{18}}.
\end{eqnarray}
 Figure~\ref{density_3} is the plot of \( P_{3}(s) \) for \( R=1 \), and
illustrates the agreement between Monte Carlo simulation and the analytical
result.

We can extend the discussion to a \( 4 \)-dimensional sphere of radius \( R \)
and arbitrary density distribution function, \begin{equation}
\rho =\rho (x_{1},x_{2},x_{3},x_{4}),
\end{equation}
 where \begin{equation}
x_{1}^{2}+x_{2}^{2}+x_{3}^{2}+x_{4}^{2}\leq R^{2}.
\end{equation}
 The \( 4 \)-dimensional hyperspherical coordinates~\cite{Luban} that are
a generalization of the conventional \( 3 \)-dimensional spherical coordinates
are defined as follows: \begin{eqnarray}
x_{1} & = & r\sin \theta _{2}\sin \theta _{1}\cos \phi ,\nonumber \\
x_{2} & = & r\sin \theta _{2}\sin \theta _{1}\sin \phi ,\nonumber \\
x_{3} & = & r\sin \theta _{2}\cos \theta _{1},\nonumber \\
x_{4} & = & r\cos \theta _{2},
\end{eqnarray}
 and \begin{eqnarray}
r & = & \sqrt{x_{1}^{2}+x_{2}^{2}+x_{3}^{2}+x_{4}^{2}},\nonumber \\
\theta _{1} & = & \tan ^{-1}\frac{\sqrt{x_{1}^{2}+x_{2}^{2}}}{x_{3}},\nonumber \\
\theta _{2} & = & \tan ^{-1}\frac{\sqrt{x_{1}^{2}+x_{2}^{2}+x_{3}^{2}}}{x_{4}},\nonumber \\
\phi  & = & \tan ^{-1}\frac{x_{2}}{x_{1}},
\end{eqnarray}
where \begin{equation}
0\leq r\leq R,\, \, \, \, \, 0\leq \theta _{1},\theta _{2}\leq \pi ,\, \, \, \, \, 0\leq \phi \leq 2\pi ,
\end{equation}
 and the volume element dV is given by \begin{equation}
dV=dx_{1}dx_{2}dx_{3}dx_{4}=r^{3}\sin ^{2}\theta _{2}\sin \theta _{1}drd\theta _{2}d\theta _{1}d\phi .
\end{equation}
 The representation of the rotation operator \( \mathbf{R}(\vec{s}\, ) \) for
a given \( \vec{s} \) is a \( 4 \)-dimensional rotation matrix \begin{eqnarray}
R_{4\times 4}(\theta _{2},\theta _{1},\phi ) & = & R_{4\times 4}(\theta _{2})\times R_{4\times 4}(\theta _{1})\times R_{4\times 4}(\phi )\nonumber \\
 & = & \left[ \begin{array}{cccc}
\cos \theta _{1}\cos \phi \,  & \, \cos \theta _{1}\sin \phi \,  & \, -\sin \theta \,  & \, 0\\
-\sin \phi \,  & \, \cos \phi \,  & \, 0\,  & \, 0\\
\cos \theta _{2}\sin \theta _{1}\cos \phi \,  & \, \cos \theta _{2}\sin \theta _{1}\sin \phi \,  & \, \cos \theta _{2}\cos \theta _{1}\,  & \, -\sin \theta _{2}\\
\sin \theta _{2}\sin \theta _{1}\cos \phi \,  & \, \sin \theta _{2}\sin \theta _{1}\sin \phi \,  & \, \sin \theta _{2}\cos \theta _{1}\,  & \, \cos \theta _{2}
\end{array}\right] ,
\end{eqnarray}
 where 

\begin{eqnarray}
R_{4\times 4}(\theta _{2}) & = & \left[ \begin{array}{cccc}
1\,  & \, 0\,  & \, 0\,  & \, 0\\
0\,  & \, 1\,  & \, 0\,  & \, 0\\
0\,  & \, 0\,  & \, \cos \theta _{2}\,  & \, -\sin \theta _{2}\\
0\,  & \, 0\,  & \, \sin \theta _{2}\,  & \, \cos \theta _{2}
\end{array}\right] ,\\
R_{4\times 4}(\theta _{1}) & = & \left[ \begin{array}{cccc}
\cos \theta _{1}\,  & \, 0\,  & \, -\sin \theta _{1}\,  & \, 0\\
0\,  & \, 1\,  & \, 0\,  & \, 0\\
\sin \theta _{1}\,  & \, 0\,  & \, \cos \theta _{1}\,  & \, 0\\
0\,  & \, 0\,  & \, 0\,  & \, 1
\end{array}\right] ,\\
R_{4\times 4}(\phi ) & = & \left[ \begin{array}{cccc}
\cos \phi \,  & \, \sin \phi \,  & \, 0\,  & \, 0\\
-\sin \phi \,  & \, \cos \phi \,  & \, 0\,  & \, 0\\
0\,  & \, 0\,  & \, 1\,  & \, 0\\
0\,  & \, 0\,  & \, 0\,  & \, 1
\end{array}\right] .
\end{eqnarray}
 It is convenient to introduce the notations, \begin{eqnarray}
\vec{\mathbf{X}}' & = & R_{4\times 4}^{T}(\theta _{2},\theta _{1},\phi )\vec{\mathbf{X}},\nonumber \\
 & = & R_{4\times 4}^{T}(\phi )R_{4\times 4}^{T}(\theta _{1})R_{4\times 4}^{T}(\theta _{2})\vec{\mathbf{X}},
\end{eqnarray}
 \begin{eqnarray}
\vec{\mathbf{X}}'' & = & R_{4\times 4}^{T}(\theta _{2},\theta _{1},\phi )(\vec{\mathbf{X}}-\vec{\mathbf{S}}),\nonumber \\
 & = & R_{4\times 4}^{T}(\phi )R_{4\times 4}^{T}(\theta _{1})R_{4\times 4}^{T}(\theta _{2})(\vec{\mathbf{X}}-\vec{\mathbf{S}}),
\end{eqnarray}
where\begin{equation}
\vec{\mathbf{X}}=\left[ \begin{array}{c}
x_{1}\\
x_{2}\\
x_{3}\\
x_{4}
\end{array}\right] ,\, \, \, \vec{\mathbf{X}}'=\left[ \begin{array}{c}
x_{1}'\\
x_{2}'\\
x_{3}'\\
x_{4}'
\end{array}\right] ,\, \, \, \vec{\mathbf{X}}''=\left[ \begin{array}{c}
x_{1}''\\
x_{2}''\\
x_{3}''\\
x_{4}''
\end{array}\right] ,\, \, \, \vec{\mathbf{S}}=\left[ \begin{array}{c}
0\\
0\\
0\\
s
\end{array}\right] .
\end{equation}
 The PDF \( P_{4}(s) \) can then be written as \begin{equation}
P_{4}(s)=\frac{s^{3}\int \left[ \theta _{2},\theta _{1},\phi \right] \times \int \left[ x_{1},x_{2},x_{3},x_{4}\right] \times \rho (\mathbf{X}')\times \rho (\mathbf{X}'')}{\int _{0}^{2R}\left\{ s^{3}\int \left[ \theta _{2},\theta _{1},\phi \right] \times \int \left[ x_{1},x_{2},x_{3},x_{4}\right] \times \rho (\mathbf{X}')\times \rho (\mathbf{X}'')\right\} ds},
\end{equation}
 where

\begin{eqnarray}
\int \left[ \theta _{2},\theta _{1},\phi \right]  & \equiv  & \int _{0}^{\pi }\sin ^{2}\theta _{2}d\theta _{2}\int _{0}^{\pi }\sin \theta _{1}d\theta _{1}\int _{0}^{2\pi }d\phi ,\\
\int \left[ x_{1},x_{2},x_{3},x_{4}\right]  & \equiv  & \int _{\frac{s}{2}}^{R}dx_{4}\int _{-\sqrt{R^{2}-x_{4}^{2}}}^{\sqrt{R^{2}-x_{4}^{2}}}dx_{1}\int _{-\sqrt{R^{2}-x_{4}^{2}-x_{1}^{2}}}^{\sqrt{R^{2}-x_{4}^{2}-x_{1}^{2}}}dx_{2}\int _{-\sqrt{R^{2}-x_{4}^{2}-x_{1}^{2}-x_{2}^{2}}}^{\sqrt{R^{2}-x_{4}^{2}-x_{1}^{2}-x_{2}^{2}}}dx_{3,}
\end{eqnarray}
 \begin{eqnarray}
\rho (\mathbf{X}') & = & \rho (x_{1}',x_{2}',x_{3}',x_{4}'),\nonumber \\
x_{1}' & = & \cos \theta _{1}\cos \phi x_{1}-\sin \phi x_{2}+\cos \theta _{2}\sin \theta _{1}\cos \phi x_{3}+\sin \theta _{2}\sin \theta _{1}\cos \phi x_{4},\nonumber \\
x_{2}' & = & \cos \theta _{1}\sin \phi x_{1}+\cos \phi x_{2}+\cos \theta _{2}\sin \theta _{1}\sin \phi x_{3}+\sin \theta _{2}\sin \theta _{1}\sin \phi x_{4},\nonumber \\
x_{3}' & = & -\sin \theta _{1}x_{1}+\cos \theta _{2}\cos \theta _{1}x_{3}+\sin \theta _{2}\cos \theta _{1}x_{4},\nonumber \\
x_{4}' & = & -\sin \theta _{2}x_{3}+\cos \theta _{2}x_{4},
\end{eqnarray}
 \begin{eqnarray}
\rho (\mathbf{X}'') & = & \rho (x_{1}'',x_{2}'',x_{3}'',x_{4}''),\nonumber \\
x_{1}'' & = & \cos \theta _{1}\cos \phi x_{1}-\sin \phi x_{2}+\cos \theta _{2}\sin \theta _{1}\cos \phi x_{3}+\sin \theta _{2}\sin \theta _{1}\cos \phi (x_{4}-s),\nonumber \\
x_{2}'' & = & \cos \theta _{1}\sin \phi x_{1}+\cos \phi x_{2}+\cos \theta _{2}\sin \theta _{1}\sin \phi x_{3}+\sin \theta _{2}\sin \theta _{1}\sin \phi (x_{4}-s),\nonumber \\
x_{3}'' & = & -\sin \theta _{1}x_{1}+\cos \theta _{2}\cos \theta _{1}x_{3}+\sin \theta _{2}\cos \theta _{1}(x_{4}-s),\nonumber \\
x_{4}'' & = & -\sin \theta _{2}x_{3}+\cos \theta _{2}(x_{4}-s).
\end{eqnarray}

\begin{figure}
\phantom{aaa}
\vfill
{\par\centering \resizebox*{6in}{!}{\includegraphics{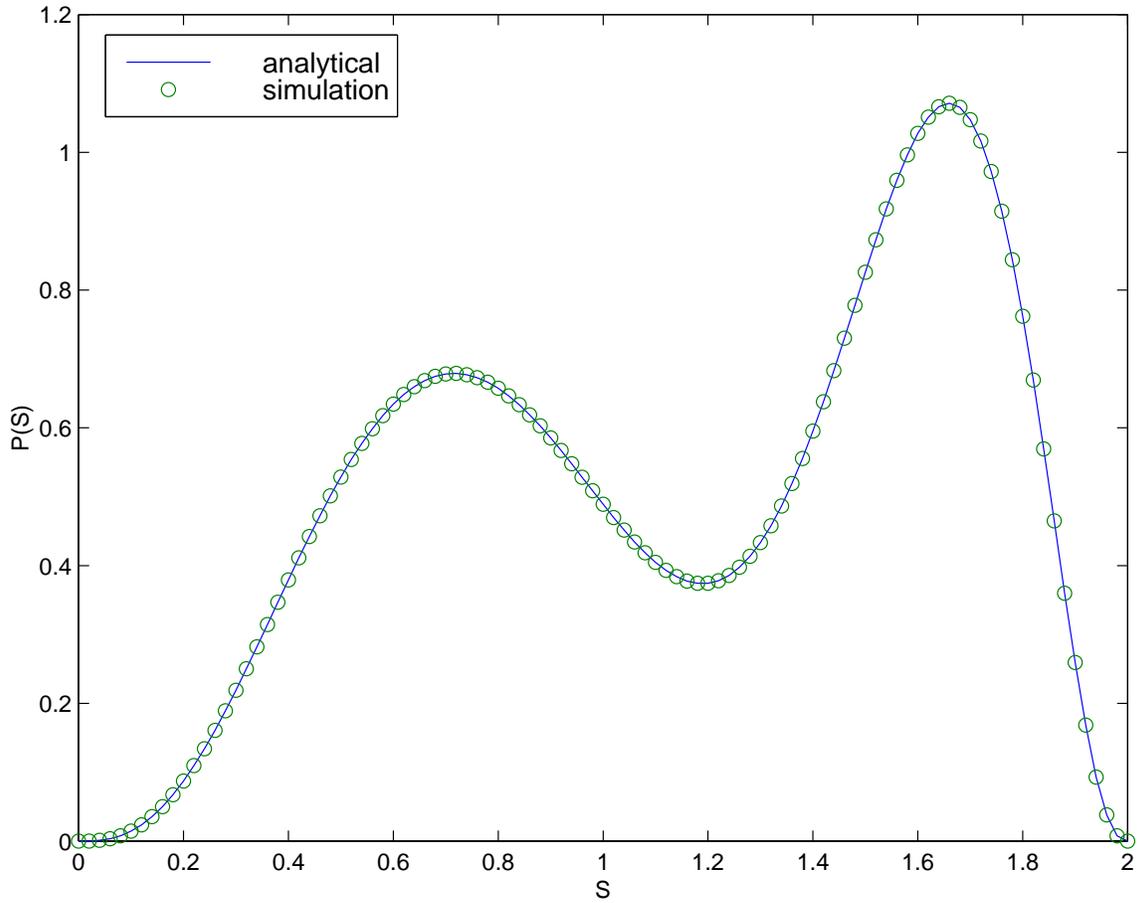}} \par}
\caption{Plot of \protect\( P_{4}(s)\protect \) as a function of \protect\( s\protect \)
for a \protect\( 4\protect \)-dimensional hypersphere of radius \protect\( R=1\protect \)
and \protect\( \rho =\left( 32N/\pi ^{2}\right) x_{1}^{4}\protect \).\label{arbitrary_4d}}
\vfill\phantom{aaa}\end{figure}

As an example, consider a \( 4 \)-dimensional sphere of radius \( R \) and
non-uniform density distribution\begin{equation}
\label{eq_example_density_4d}
\rho (x_{1},x_{2},x_{3},x_{4})=\frac{32N}{\pi ^{2}R^{8}}x_{1}^{4}=\frac{12N}{\pi ^{2}R^{6}}r^{4}\sin ^{4}\theta _{2}\sin ^{4}\theta _{1}\cos ^{4}\phi ,
\end{equation}
 where the normalizing factor \( N \) is given by \begin{equation}
N=\int _{-R}^{R}dx_{1}\int _{-\sqrt{R^{2}-x_{1}^{2}}}^{\sqrt{R^{2}-x_{1}^{2}}}dx_{2}\int _{-\sqrt{R^{2}-x_{1}^{2}-x_{2}^{2}}}^{\sqrt{R^{2}-x_{1}^{2}-x_{2}^{2}}}dx_{3}\int _{-\sqrt{R^{2}-x_{1}^{2}-x_{2}^{2}-x_{3}^{2}}}^{\sqrt{R^{2}-x_{1}^{2}-x_{2}^{2}-x_{3}^{2}}}\rho (x_{1},x_{2},x_{3},x_{4})dx_{4}.
\end{equation}
 Then\begin{eqnarray}
P_{4}(s) & = & \frac{56}{3}\frac{s^{3}}{R^{4}}+48\frac{s^{5}}{R^{6}}+8\frac{s^{7}}{R^{8}}\nonumber \\
 &  & -\frac{1}{\pi }\left( \frac{196}{3}\frac{s^{4}}{R^{6}}+\frac{114}{5}\frac{s^{6}}{R^{8}}+\frac{28}{15}\frac{s^{8}}{R^{10}}-\frac{4}{5}\frac{s^{10}}{R^{12}}+\frac{2}{9}\frac{s^{12}}{R^{14}}-\frac{1}{45}\frac{s^{14}}{R^{16}}\right) \times \sqrt{4R^{2}-s^{2}}\nonumber \\
 &  & -\frac{1}{\pi }\left( \frac{112}{3}\frac{s^{3}}{R^{4}}+96\frac{s^{5}}{R^{6}}+16\frac{s^{7}}{R^{8}}\right) \times \sin ^{-1}\left( \frac{s}{2R}\right) .
\end{eqnarray}
 Figure~\ref{arbitrary_4d} is the plot of \( P_{4}(s) \) for \( R=1 \), and
illustrates the agreement between the Monte Carlo simulation and the analytical
result.

We turn next to the general case of an \( n \)-dimensional sphere of radius
\( R \) and arbitrary density distribution,\begin{equation}
\rho =\rho (x_{1},x_{2},\cdots ,x_{n}),
\end{equation}
 where \begin{equation}
x_{1}^{2}+x_{2}^{2}+\cdots +x_{n}^{2}\leq R^{2}.
\end{equation}
 Define the following \( n \)-dimensional spherical coordinates \( x_{1},\, \ldots ,\, x_{n} \)~\cite{fleming},
\begin{eqnarray}
x_{1} & = & r\sin \theta _{n-2}\sin \theta _{n-3}\cdots \cdots \cdots \cdots \cdots \cdots \sin \theta _{2}\sin \theta _{1}\cos \phi ,\nonumber \\
x_{2} & = & r\sin \theta _{n-2}\sin \theta _{n-3}\cdots \cdots \cdots \cdots \cdots \cdots \sin \theta _{2}\sin \theta _{1}\sin \phi ,\nonumber \\
x_{3} & = & r\sin \theta _{n-2}\sin \theta _{n-3}\cdots \cdots \cdots \cdots \cdots \cdots \sin \theta _{2}\cos \theta _{1},\nonumber \\
\vdots  & \vdots  & \vdots \nonumber \\
x_{i} & = & r\sin \theta _{n-2}\sin \theta _{n-3}\cdots \cdots \sin \theta _{i-1}\cos \theta _{i-2},\nonumber \\
\vdots  & \vdots  & \vdots \nonumber \\
x_{n-2} & = & r\sin \theta _{n-2}\sin \theta _{n-3}\cos \theta _{n-4},\nonumber \\
x_{n-1} & = & r\sin \theta _{n-2}\cos \theta _{n-3},\nonumber \\
x_{n} & = & r\cos \theta _{n-2},
\end{eqnarray}
 where\begin{eqnarray}
dV & = & dx_{1}dx_{2}\cdots dx_{n}\nonumber \\
 & = & r^{n-1}\sin ^{n-2}\theta _{n-2}\sin ^{n-3}\theta _{n-3}\cdots \cdots \sin ^{2}\theta _{2}\sin \theta _{1}\, drd\theta _{n-2}d\theta _{n-3}\cdots \cdots d\theta _{2}d\theta _{1}d\phi ,
\end{eqnarray}
 and \begin{equation}
0\leq r\leq R,\, \, \, \, \, 0\leq \theta _{1},\theta _{2},\cdots \cdots ,\theta _{n-3},\theta _{n-2}\leq \pi ,\, \, \, \, \, 0\leq \phi \leq 2\pi .
\end{equation}
 The rotation operator \( \mathbf{R}(\vec{s}\, ) \) for a given \( \vec{s} \)
is \begin{equation}
\label{eq_n_rotation_operator}
R_{n\times n}(\theta _{n-2},\theta _{n-3},\cdots \theta _{2},\theta _{1},\phi )=R_{n\times n}(\theta _{n-2})R_{n\times n}(\theta _{n-3})\cdots R_{n\times n}(\theta _{2})R_{n\times n}(\theta _{1})R_{n\times n}(\phi ).
\end{equation}
 The matrix \( R_{n\times n}(\phi ) \) appearing in Eq.~(\ref{eq_n_rotation_operator})
has the following elements: \( R_{11}(\phi )=\cos \phi  \), \( R_{12}(\phi )=\sin \phi  \),
\( R_{21}(\phi )=-\sin \phi  \) , \( R_{22}(\phi )=\cos \phi  \) , and \( R_{lm}(\phi )=\delta _{lm} \)
where \( l,m\neq 1,2 \) and \( 1\leq l,m\leq n \). The matrix \( R_{n\times n}(\theta _{1}) \)
has the following elements: \( R_{11}(\theta _{1})=\cos \theta _{1} \), \( R_{13}(\theta _{1})=-\sin \theta _{1} \),
\( R_{31}(\theta _{1})=\sin \theta _{1} \), \( R_{33}(\theta _{1})=\cos \theta _{1} \),
and \( R_{lm}(\theta _{1})=\delta _{lm} \) where \( l,m\neq 1,3 \) and \( 1\leq l,m\leq n \).
The matrix \( R_{n\times n}(\theta _{i}) \) has the following elements: \( R_{i+1,i+1}(\theta _{i})=\cos \theta _{i} \),
\( R_{i+1,i+2}(\theta _{i})=-\sin \theta _{i} \), \( R_{i+2,i+1}(\theta _{i})=\sin \theta _{i} \),
\( R_{i+2,i+2}(\theta _{i})=\cos \theta _{i} \), and \( R_{lm}(\theta _{i})=\delta _{lm} \)
where \( 2\leq i\leq n-2 \), \( l,m\neq i+1,i+2 \), and \( 1\leq l,m\leq n \).
Notice that \( R_{n\times n}(\theta _{n-2},\theta _{n-3},\cdots \theta _{2},\theta _{1},\phi ) \)
has the following properties: 

\begin{enumerate}
\item \( R_{n\times n}(\theta _{n-2},\theta _{n-3},\cdots \theta _{2},\theta _{1},\phi ) \)
is an orthogonal matrix such that \( R_{n\times n}^{-1}=R_{n\times n}^{T} \).
\item The \( 1 \)st row matrix elements are \( R_{11}=\cos \theta _{1}\cos \phi  \),
\( R_{12}=\cos \theta _{1}\sin \phi  \), \( R_{13}=-\sin \theta _{1} \), and
\( R_{1j}=0 \) for \( 4\leq j\leq n \).
\item The \( 2 \)nd row matrix elements are \( R_{21}=-\sin \phi  \), \( R_{22}=\cos \phi  \),
and \( R_{2j}=0 \) for \( 3\leq j\leq n \).
\item The \( i \)th row matrix elements, where \( 3\leq i\leq n-1 \), are \( R_{i1}=\cos \theta _{i-1}\times x_{1}[i] \),
\( R_{i2}=\cos \theta _{i-1}\times x_{2}[i] \), \( R_{im}=\cos \theta _{i-1}\times x_{m}[i] \)
for \( 1\leq m\leq i \), \( R_{ii}=\cos \theta _{i-1}\times x_{i}[i] \), \( R_{i\, i+1}=-\sin \theta _{i-1} \),
and \( R_{ij}=0 \) for \( i+2\leq j\leq n \), where \( x_{m}[i] \) is the
\( m \)th component of the \( i \)-dimensional Cartesian coordinate system
in the representation of the \( i \)-dimensional spherical coordinate system
for a unit vector. Some examples are \( x_{3}[3]=\cos \theta _{1} \), \( x_{3}[4]=\sin \theta _{2}\cos \theta _{1} \),
\( x_{3}[5]=\sin \theta _{3}\sin \theta _{2}\cos \theta _{1} \), and \( x_{3}[6]=\sin \theta _{4}\sin \theta _{3}\sin \theta _{2}\cos \theta _{1} \).
\item The \( n \)th row matrix elements are \( R_{nj}=x_{j}[n] \) for \( 1\leq j\leq n \),
where \( x_{j}[n] \) is the \( j \)th component of the \( n \)-dimensional
Cartesian coordinate in the representation of the \( n \)-dimensional spherical
coordinate system for a unit vector. Some examples are \begin{eqnarray}
x_{1}[n] & = & \sin \theta _{n-2}\sin \theta _{n-3}\cdots \cdots \sin \theta _{2}\sin \theta _{1}\cos \phi ,\nonumber \\
x_{2}[n] & = & \sin \theta _{n-2}\sin \theta _{n-3}\cdots \cdots \sin \theta _{2}\sin \theta _{1}\sin \phi ,\nonumber \\
x_{3}[n] & = & \sin \theta _{n-2}\sin \theta _{n-3}\cdots \cdots \sin \theta _{2}\cos \theta _{1},\nonumber \\
x_{n}[n] & = & \cos \theta _{n-2.}
\end{eqnarray}

\end{enumerate}

The final master probability density function formula \( P_{n}(s) \) for an
\( n \)-dimensional sphere of radius \( R \) and arbitrary density distribution
has the following mathematical representation: \begin{equation}
\label{eq_n_master_founula}
P_{n}(s)=\frac{s^{n-1}\int [\theta _{n-2},\theta _{n-3},\cdots \cdots \theta _{2},\theta _{1},\phi ]\int [x_{1},x_{2},\cdots \cdots ,x_{n}]\rho (\mathbf{X}')\rho (\mathbf{X}'')}{\int _{0}^{2R}\left\{ s^{n-1}\int [\theta _{n-2},\theta _{n-3},\cdots \cdots \theta _{2},\theta _{1},\phi ]\int [x_{1},x_{2},\cdots \cdots ,x_{n}]\rho (\mathbf{X}')\rho (\mathbf{X}'')\right\} ds},
\end{equation}
 where \begin{eqnarray}
\int [\theta _{n-2},\theta _{n-3},\cdots \cdots \theta _{2},\theta _{1},\phi ] & \equiv  & \int _{0}^{\pi }\sin ^{n-2}\theta _{n-2}d\theta _{n-2}\int _{0}^{\pi }\sin ^{n-3}\theta _{n-3}d\theta _{n-3}\cdots \cdots \nonumber \\
 &  & \cdots \cdots \int _{0}^{\pi }\sin ^{2}\theta _{2}d\theta _{2}\int _{0}^{\pi }\sin \theta _{1}d\theta _{1}\int _{0}^{2\pi }d\phi ,
\end{eqnarray}
 \begin{eqnarray}
\int [x_{1},x_{2},\cdots \cdots ,x_{n}] & \equiv  & \int _{\frac{s}{2}}^{R}dx_{n}\int _{-\sqrt{R^{2}-x_{n}^{2}}}^{\sqrt{R^{2}-x_{n}^{2}}}dx_{1}\int _{-\sqrt{R^{2}-x_{n}^{2}-x_{1}^{2}}}^{\sqrt{R^{2}-x_{n}^{2}-x_{1}^{2}}}dx_{2}\cdots \cdots \nonumber \\
 &  & \cdots \cdots \int _{-\sqrt{R^{2}-x_{n}^{2}-x_{1}^{2}-\cdots x_{n-3}^{2}}}^{\sqrt{R^{2}-x_{n}^{2}-x_{1}^{2}-\cdots x_{n-3}^{2}}}dx_{n-2}\int ^{\sqrt{R^{2}-x_{n}^{2}-x_{1}^{2}-\cdots x_{n-2}^{2}}}_{-\sqrt{R^{2}-x_{n}^{2}-x_{1}^{2}-\cdots x_{n-2}^{2}}}dx_{n-1},
\end{eqnarray}
\begin{eqnarray}
\rho (\mathbf{X}') & = & \rho (x_{1}',x_{2}',\cdots \cdots x_{n}'),\\
\rho (\mathbf{X}'') & = & \rho (x_{1}'',x_{2}'',\cdots \cdots x_{n}'').
\end{eqnarray}
 Additionally,  we introduce the following notations: \begin{eqnarray}
\vec{\mathbf{X}}' & = & R_{n\times n}^{T}(\theta _{n-2},\theta _{n-3},\cdots \cdots \theta _{2},\theta _{1},\phi )\vec{\mathbf{X}}\nonumber \\
 & = & R_{n\times n}^{T}(\phi )R_{n\times n}^{T}(\theta _{1})R_{n\times n}^{T}(\theta _{2})\cdots \cdots R_{n\times n}^{T}(\theta _{n-3})R_{n\times n}^{T}(\theta _{n-2})\vec{\mathbf{X}},\\
\vec{\mathbf{X}}'' & = & R_{n\times n}^{T}(\theta _{n-2},\theta _{n-3},\cdots \cdots \theta _{2},\theta _{1},\phi )(\vec{\mathbf{X}}-\vec{\mathbf{S}})\nonumber \\
 & = & R_{n\times n}^{T}(\phi )R_{n\times n}^{T}(\theta _{1})R_{n\times n}^{T}(\theta _{2})\cdots \cdots R_{n\times n}^{T}(\theta _{n-3})R_{n\times n}^{T}(\theta _{n-2})(\vec{\mathbf{X}}-\vec{\mathbf{S}}),
\end{eqnarray}
 where \begin{equation}
\vec{\mathbf{X}}=\left[ \begin{array}{c}
x_{1}\\
x_{2}\\
\vdots \\
\vdots \\
x_{n-1}\\
x_{n}
\end{array}\right] ,\, \, \, \vec{\mathbf{X}}'=\left[ \begin{array}{c}
x_{1}'\\
x_{2}'\\
\vdots \\
\vdots \\
x_{n-1}'\\
x_{n}'
\end{array}\right] ,\, \, \, \vec{\mathbf{X}}''=\left[ \begin{array}{c}
x_{1}''\\
x_{2}''\\
\vdots \\
\vdots \\
x_{n-1}''\\
x_{n}''
\end{array}\right] ,\, \, \, \vec{\mathbf{S}}=\left[ \begin{array}{c}
0\\
0\\
\vdots \\
\vdots \\
0\\
s
\end{array}\right] .
\end{equation}

The technique for generating random points within an \( n \)-dimensional sphere
having an arbitrary density distribution will be discussed elsewhere~\cite{sjtu,sjtu2}.

\section{Applications}
\label{applications}

\subsection{\protect\( m\protect \)th moment}

We first calculate the \( m \)th moment \( \left\langle s^{m}\right\rangle  \)
for the case of \( n \)-dimensional uniform sphere, where \begin{equation}
\left\langle s^{m}\right\rangle =\int _{0}^{2R}s^{m}P_{n}(s)ds,
\end{equation}
 and \( m \) is a positive integer. Evidently \( \left\langle s^{m}\right\rangle  \)
gives the expectation (average) value of the \( m \)th power of the distance
between two independent random points generated inside a uniform \( n \)-dimensional
sphere. By utilizing the function \( C\left( a;m,n\right)  \) defined in Eq.~(\ref{eq_notation_001}),
we can write \( \left\langle s^{m}\right\rangle  \) as\begin{equation}
\label{eq_moment_001}
\left\langle s^{m}\right\rangle =\frac{C(2R;m,n)}{C(2R;0,n)}=2^{m+n}\left( \frac{n}{m+n}\right) \frac{B\left( \frac{n}{2}+\frac{1}{2},\frac{n}{2}+\frac{1}{2}+\frac{m}{2}\right) }{B\left( \frac{n}{2}+\frac{1}{2},\frac{1}{2}\right) }R^{m},
\end{equation}
 where we have used the following identity: \begin{eqnarray}
C(a;m,n) & = & \frac{1}{2}\frac{a^{m+n}}{m+n}B\left( \frac{1}{2},\frac{n}{2}+\frac{1}{2}\right) R^{n}-\frac{1}{2}\frac{a^{m+n}}{m+n}B_{\left( \frac{a}{2R}\right) ^{2}}\left( \frac{1}{2},\frac{n}{2}+\frac{1}{2}\right) R^{n}\nonumber \\
 &  & +\frac{1}{2}\frac{(2R)^{m+n}}{m+n}B_{\left( \frac{a}{2R}\right) ^{2}}\left( \frac{n}{2}+\frac{1}{2}+\frac{m}{2},\frac{n}{2}+\frac{1}{2}\right) R^{n}.
\end{eqnarray}
 Furthermore we can rewrite Eq.~(\ref{eq_moment_001}) in terms of the gamma
function by replacing the beta functions as follows: \begin{eqnarray}
\left\langle s^{m}\right\rangle  & = & (2R)^{m}\left( \frac{n}{m+n}\right) \frac{\Gamma \left( \frac{n}{2}+\frac{m}{2}+\frac{1}{2}\right) \Gamma \left( n+1\right) }{\Gamma \left( n+1+\frac{m}{2}\right) \Gamma \left( \frac{n}{2}+\frac{1}{2}\right) }\label{eq_moment_1} \\
 & = & \left( \frac{n}{m+n}\right) ^{2}\frac{\Gamma \left( n+m+1\right) \Gamma \left( \frac{n}{2}\right) }{\Gamma \left( \frac{n}{2}+\frac{m}{2}\right) \Gamma \left( n+1+\frac{m}{2}\right) }R^{m}.\label{eq_moment_2} 
\end{eqnarray}
 The results in Eqs.~(\ref{eq_moment_1}) and (\ref{eq_moment_2}) are identical
to those given in Refs.~\cite{Kendall,Santalo}. They can be extended to evaluate
\( \left\langle s^{-m}\right\rangle  \) and we find \begin{equation}
\label{eq_moment_002}
\left\langle \frac{1}{s^{m}}\right\rangle =\frac{n}{n-m}\frac{2^{n-m}}{R^{m}}\frac{B\left( \frac{n}{2}+\frac{1}{2},\frac{n}{2}+\frac{1}{2}-\frac{m}{2}\right) }{B\left( \frac{n}{2}+\frac{1}{2},\frac{1}{2}\right) },
\end{equation}
 where \begin{equation}
m\leq n-1.
\end{equation}
 Combining Eqs.~(\ref{eq_moment_001}) and (\ref{eq_moment_002}), the \( m \)th
moment \( \left\langle s^{m}\right\rangle  \) has the general form \begin{equation}
\label{eq_moment_uniform_tu}
\left\langle s^{m}\right\rangle =2^{n+m}\left( \frac{n}{n+m}\right) \frac{B\left( \frac{n}{2}+\frac{1}{2},\frac{n}{2}+\frac{1}{2}+\frac{m}{2}\right) }{B\left( \frac{n}{2}+\frac{1}{2},\frac{1}{2}\right) }R^{m},
\end{equation}
 where \begin{equation}
m=-(n-1),\, -(n-2)\, ,\ldots ,\, -2,\, -1,\, 0,\, 1,\, 2,\, \ldots \ldots .
\end{equation}
 Following is a short list of \( \left\langle s^{m}\right\rangle  \) in \( 3 \)
dimensions: \( \left\langle 1/s^{2}\right\rangle  \)=\( 9/4 \), \( \left\langle 1/s\right\rangle  \)=\( 6/5 \),
\( \left\langle s\right\rangle  \)=\( 36/35 \), \( \left\langle s^{2}\right\rangle  \)=\( 6/5 \),
\( \left\langle s^{3}\right\rangle  \)=\( 32/21 \), \( \left\langle s^{4}\right\rangle  \)=\( 72/35 \),
and \( \left\langle s^{5}\right\rangle  \)=\( 32/11 \), where the radius \( R \)
has been set to unity. 

Additionally, \( \left\langle s^{m}\right\rangle  \) can be evaluated for a
sphere having a Gaussian density distribution and radius \( R\rightarrow \infty  \).
From Eq.~(\ref{eq_n_pdf_gaussian}) we have,\begin{equation}
\label{eq_moment_gaussian}
\left\langle s^{m}\right\rangle =\lim _{R\rightarrow \infty }\int _{0}^{2R}s^{m}P_{n}(s)ds=(2\sigma )^{m}\frac{\Gamma \left( \frac{n+m}{2}\right) }{\Gamma \left( \frac{n}{2}\right) }.
\end{equation}

In some applications involving low-energy interactions among nucleons the lower
limit (zero) should be replaced by the hard-core radius \( r_{c}\cong 0.5\times 10^{-13} \)
cm~\cite{Bohr}. In such cases the expressions for \( P_{n}(s) \) and \( \left\langle s^{m}\right\rangle  \)
assume the form:\begin{equation}
P_{n}(s)=\frac{s^{n-1}\int _{\frac{s}{2}}^{R}\left( R^{2}-x^{2}\right) ^{\frac{n-1}{2}}dx}{\int _{r_{c}}^{2R}ds\int _{\frac{s}{2}}^{R}dx\, s^{n-1}\left( R^{2}-x^{2}\right) ^{\frac{n-1}{2}}}=\frac{s^{n-1}\int _{\frac{s}{2}}^{R}\left( R^{2}-x^{2}\right) ^{\frac{n-1}{2}}dx}{C(2R;0,n)-C(r_{c};0,n)}
\end{equation}
 and\begin{equation}
\left\langle s^{m}\right\rangle =\int _{r_{c}}^{2R}s^{m}P_{n}(s)ds=\frac{H(R,\, r_{c;}\, m,\, n)}{H(R,\, r_{c};\, 0,\, n)},
\end{equation}
where \begin{eqnarray}
H\left( R,\, r_{c};\, m,\, n\right)  & = & \frac{(2R)^{n+m}}{n+m}\left[ B\left( \frac{n}{2}+\frac{1}{2},\frac{n}{2}+\frac{1}{2}+\frac{m}{2}\right) -B_{\left( \frac{r_{c}}{2R}\right) ^{2}}\left( \frac{n}{2}+\frac{1}{2},\frac{n}{2}+\frac{1}{2}+\frac{m}{2}\right) \right] \nonumber \\
 &  & -\frac{r_{c}^{n+m}}{n+m}\left[ B\left( \frac{1}{2},\frac{n}{2}+\frac{1}{2}\right) -B_{\left( \frac{r_{c}}{2R}\right) ^{2}}\left( \frac{1}{2},\frac{n}{2}+\frac{1}{2}\right) \right] ,\label{eq_moment_bohr} 
\end{eqnarray}
and \( m \) is an integer.

\subsection{Coulomb Self-Energy of a Collection of Charges}
\label{section_coulomb}

As an application of the preceding formalism we evaluate the electrostatic energy
\( W_{n} \) of a collection of \( Z \) charges in \( n \) dimensions by applying
geometric probability techniques. Consider first the familiar case of a spherical
charge distribution in \( 3 \) dimensions. For each pair of charges the potential
energy due to the Coulomb interaction in Gaussian units is \begin{equation}
\label{eq_coulomb}
V\left( \left| \vec{r}_{2}-\vec{r}_{1}\right| \right) =\frac{e_{0}^{2}}{\left| \vec{r}_{2}-\vec{r}_{1}\right| },
\end{equation}
where \( e_{0} \) is the elementary charge (\( e_{0}^{2}/\hbar c\cong 1/137 \)).
Hence if we assume that the charges in each pair are uniformly distributed within
the same spherical volume of radius \( R \) then the average Coulomb energy
\( U_{3} \) of each pair of charges is\begin{equation}
\label{eq_coulomv_u3}
U_{3}=e^{2}\int _{0}^{2R}\frac{1}{s}P_{3}(s)ds=\frac{6}{5}\frac{e_{0}^{2}}{R}.
\end{equation}
 For a collection of \( Z \) charges there are \( Z(Z-1)/2 \) such pairs,
and hence the total Coulomb energy \( W_{3} \) is \begin{equation}
\label{eq_coulomb_w3}
W_{3}=\frac{Z(Z-1)}{2}U_{3}=\frac{3}{5}Z(Z-1)\frac{e_{0}^{2}}{R}.
\end{equation}
 For \( n\geq 3 \) the Coulomb potential energy between two charges has the
general form~\cite{Burgbacher} \begin{equation}
\label{eq_n_coulomb}
V_{n}\left( \left| \vec{r}_{2}-\vec{r}_{1}\right| \right) =\frac{q_{n}^{2}}{\left| \vec{r}_{2}-\vec{r}_{1}\right| ^{n-2}},
\end{equation}
 where \( q_{n} \) is a suitably defined charge with appropriate dimensions.
Hence Eq.~(\ref{eq_coulomv_u3}) generalizes to\begin{equation}
\label{eq_coulomb_u_n}
U_{n}=q_{n}^{2}\int _{0}^{2R}\frac{1}{s^{n-2}}P_{n}(s)ds=q_{n}^{2}\left\langle \frac{1}{s^{n-2}}\right\rangle .
\end{equation}
 Using the results of Eq.~(\ref{eq_moment_uniform_tu}) with \( m=n-2 \) we
then find\begin{equation}
W_{n}=Z(Z-1)\frac{B\left( \frac{n}{2}+\frac{1}{2},\, \frac{3}{2}\right) }{B\left( \frac{n}{2}+\frac{1}{2},\, \frac{1}{2}\right) }\frac{q_{n}^{2}}{R^{n-2}}=\left( \frac{2n}{n+2}\right) \frac{Z(Z-1)q_{n}^{2}}{R^{n-2}}.
\end{equation}
 For \( n \) very large \( W_{n} \) assumes the limiting form\begin{equation}
\lim _{n\rightarrow \infty }W_{n}\cong 2\frac{Z(Z-1)q_{n}^{2}}{R^{n-2}}.
\end{equation}
 If the density distribution \( \rho _{n}(r) \) of the charges is Gaussian
rather than uniform, where \begin{equation}
\rho _{n}(r)=\frac{q_{n}}{(2\pi )^{\frac{n}{2}}\sigma _{n}^{n}}e^{-\frac{1}{2}\frac{r^{2}}{\sigma _{n}^{2}}}\, ,
\end{equation}
 then\begin{equation}
W_{n}=\frac{1}{2^{n-2}\Gamma \left( \frac{n}{2}\right) }\frac{q_{n}^{2}}{\sigma ^{n-2}}.
\end{equation}
 In the limit \( n\rightarrow \infty  \),\begin{equation}
W_{n}\cong \frac{2}{\pi }e^{n/2}n^{-n/2}\frac{q_{n}^{2}}{\sigma ^{n-2}}.
\end{equation}

These results are of interest in the context of recent work on modifications
to the Newtonian inverse-square law arising from the existence of extra spatial
dimensions~\cite{ArKani}. It is well known that the gravitational interaction
is weaker in a space with \( n>3 \) spatial dimensions, this weakness being
manifested by an inverse-power-law for the force \( F_{n} \), \( F_{n}\propto 1/r^{n-1} \).

\subsection{Neutrino-Pair Exchange Interactions}

A second example of interest is the \( \nu \bar{\nu } \)-exchange (neutrino-pair
exchange) contribution\cite{Feinberg_1,Feinberg_2,Hsu,Fischbach} to the self
energy of a nucleus or a neutron star. For two point masses the \( 2 \)-body
potential energy is given by\begin{equation}
V_{\nu \bar{\nu }}\left( \left| \vec{r}_{i}-\vec{r}_{j}\right| \right) =\frac{G_{F}^{2}\, a_{i}a_{j}}{4\pi ^{3}\left| \vec{r}_{i}-\vec{r}_{j}\right| ^{5}},
\end{equation}
 where \( a_{i} \) and \( a_{j} \) are coupling constants which characterize
the strength of the neutrino coupling to fermions \( i \) and \( j \) (\( i \),
\( j \) = electron, proton, or neutron). In the standard model~\cite{data_group},\begin{eqnarray*}
a_{e} & = & \frac{1}{2}+2\sin ^{2}\theta _{W}=0.964\\
a_{p} & = & \frac{1}{2}-2\sin ^{2}\theta _{W}=0.036\\
a_{n} & = & -\frac{1}{2}.
\end{eqnarray*}
 In contrast to the Coulomb interaction, the functional form of \( V_{\nu \bar{\nu }}(r) \)
cannot be determined in a space of arbitrary dimensions on the basis of a general
argument utilizing Gauss' law. Hence we restrict our attention here to \( 3 \)
spatial dimensions and consider the case of a sphere of radius \( R \) containing
\( N \) neutrons. For the case of a uniform density distribution we then find
\begin{equation}
\label{eq_moment_hard_core}
W_{3}=\frac{N(N-1)}{2}\left( \frac{3}{2r_{c}^{2}R^{3}}-\frac{9}{4r_{c}R^{4}}+\frac{9}{8R^{5}}-\frac{3r_{c}}{16R^{6}}\right) \frac{G_{F}^{2}}{4\pi ^{3}}\cong \frac{3N(N-1)G_{F}^{2}}{16\pi ^{3}r_{c}^{2}R^{3}},
\end{equation}
 where \( r_{c} \) is the hard-core radius. The analogous result for a Gaussian
density distribution is \begin{equation}
W_{3}=\left[ \frac{1}{r_{c}^{2}}e^{-r_{c}^{2}/4\sigma ^{2}}-\frac{\Gamma \left( 0,\, r_{c}^{2}/4\sigma ^{2}\right) }{4\sigma ^{2}}\right] \frac{N(N-1)G_{F}^{2}}{32\sigma ^{3}\pi ^{7/2}},
\end{equation}
where \( \Gamma (a,b) \) is the incomplete gamma function. The expression in
Eq.~(\ref{eq_moment_hard_core}) agrees with the result obtained in Refs.~\cite{Fischbach,Fischbach_k_t}.

\subsection{Neutron Star Models}

Another application of current interest is the self-energy of neutron star arising
from the exchange of \( \nu \bar{\nu } \) pairs. Here we evaluate the probability
density functions in \( 3 \) dimensions for neutron stars with a multiple-shell
uniform density distribution, which is what is typically assumed in neutron
star models~\cite{Pines,Shapiro}. For illustrative purposes, we discuss spherically
symmetric models with spherical \( 2 \), \( 3 \), and \( 4 \) shells, where
for simplicity we assume shells of equal thickness. Some other multiple-shell
models and their \( n \)-dimensional probability density functions can be found
in~\cite{sjtu}.

For a \( 2 \)-shell model with a uniform density in each shell define \( \rho =\rho _{1} \)
for \( 0\leq r\leq R/2 \) and \( \rho =\rho _{2} \) for \( R/2\leq r\leq R \),
where \( \rho _{1} \) and \( \rho _{2} \) are constants and \( r \) is measured
from the center of the neutron star in \( 3 \) dimensions. Using the preceding
formalism we can show that the PDF has \( 4 \) different functional forms specified
by \( 4 \) regions:

\begin{enumerate}
\item \( 0\leq s\leq \frac{1}{2}R: \)\begin{equation}
P_{3}(s)=\frac{24(\rho _{1}^{2}+7\rho _{2}^{2})s^{2}}{\left( \rho _{1}+7\rho _{2}\right) ^{2}R^{3}}-\frac{36(\rho _{1}^{2}-2\rho _{1}\rho _{2}+5\rho _{2}^{2})s^{3}}{\left( \rho _{1}+7\rho _{2}\right) ^{2}R^{4}}+\frac{12(\rho _{1}^{2}-2\rho _{1}\rho _{2}+2\rho _{2}^{2})s^{5}}{\left( \rho _{1}+7\rho _{2}\right) ^{2}R^{6}},
\end{equation}

\item \( \frac{1}{2}R\leq s\leq R: \)\begin{eqnarray}
P_{3}(s) & = & -\frac{81(\rho _{1}-\rho _{2})\rho _{2}s}{2\left( \rho _{1}+7\rho _{2}\right) ^{2}R^{2}}+\frac{24\rho _{1}s^{2}}{\left( \rho _{1}+7\rho _{2}\right) R^{3}}-\frac{36\rho _{1}(\rho _{1}+3\rho _{2})s^{3}}{\left( \rho _{1}+7\rho _{2}\right) ^{2}R^{4}}\nonumber \\
 &  & +\frac{12\rho _{1}^{2}s^{5}}{\left( \rho _{1}+7\rho _{2}\right) ^{2}R^{6}},
\end{eqnarray}

\item \( R\leq s\leq \frac{3}{2}R: \)\begin{eqnarray}
P_{3}(s) & = & -\frac{81(\rho _{1}-\rho _{2})\rho _{2}s}{2\left( \rho _{1}+7\rho _{2}\right) ^{2}R^{2}}+\frac{24(9\rho _{1}-\rho _{2})\rho _{2}s^{2}}{\left( \rho _{1}+7\rho _{2}\right) ^{2}R^{3}}-\frac{36(5\rho _{1}-\rho _{2})\rho _{2}s^{3}}{\left( \rho _{1}+7\rho _{2}\right) ^{2}R^{4}}\nonumber \\
 &  & +\frac{12(2\rho _{1}-\rho _{2})\rho _{2}s^{5}}{\left( \rho _{1}+7\rho _{2}\right) ^{2}R^{6}},
\end{eqnarray}

\item \( \frac{3}{2}R\leq s\leq 2R: \)\begin{equation}
P_{3}(s)=\frac{192\rho _{2}^{2}s^{2}}{\left( \rho _{1}+7\rho _{2}\right) ^{2}R^{3}}-\frac{144\rho _{2}^{2}s^{3}}{\left( \rho _{1}+7\rho _{2}\right) ^{2}R^{4}}+\frac{12\rho _{2}^{2}s^{5}}{\left( \rho _{1}+7\rho _{2}\right) ^{2}R^{6}}.
\end{equation}
 
\end{enumerate}
We observe that the PDFs defined in adjacent regions are continuous across the
boundaries separating the regions. 

For a \( 3 \)-shell model of a sphere of radius \( R \) in \( 3 \) dimensions,
with a different uniform density in each shell, define \( \rho =\rho _{1} \)
for \( 0\leq r\leq R/3 \), \( \rho =\rho _{2} \) for \( R/3\leq r\leq 2R/3 \),
and \( \rho =\rho _{3} \) for \( 2R/3\leq r\leq R \), where \( \rho _{1} \),
\( \rho _{2} \), and \( \rho _{3} \) are constants and \( r \) is measured
from the center of the neutron star. In this case the PDF has \( 6 \) different
functional forms specified by \( 6 \) regions:

\begin{enumerate}
\item \( 0\leq s\leq \frac{1}{3}R: \)\begin{eqnarray}
P_{3}(s) & = & \frac{81(\rho _{1}^{2}+7\rho ^{2}_{2}+19\rho ^{2}_{3})s^{2}}{(\rho _{1}+7\rho _{2}+19\rho _{3})^{2}R^{3}}-\frac{729(\rho _{1}^{2}-2\rho _{1}\rho _{2}+5\rho _{2}^{2}-8\rho _{2}\rho _{3}+13\rho _{3}^{2})s^{3}}{4(\rho _{1}+7\rho _{2}+19\rho _{3})^{2}R^{4}}\nonumber \\
 &  & +\frac{2187(\rho _{1}^{2}-2\rho _{1}\rho _{2}+2\rho _{2}^{2}-2\rho _{2}\rho _{3}+2\rho _{3}^{2})s^{5}}{16(\rho _{1}+7\rho _{2}+19\rho _{3})^{2}R^{6}},\label{eq_3_shell_001} 
\end{eqnarray}

\item \( \frac{1}{3}R\leq s\leq \frac{2}{3}R: \)\begin{eqnarray}
P_{3}(s) & = & -\frac{81(\rho _{2}-\rho _{3})(9\rho 1-9\rho _{2}+25\rho _{3})s}{8(\rho _{1}+7\rho _{2}+19\rho _{3})^{2}R^{2}}+\frac{81(\rho _{1}^{2}+7\rho _{1}\rho _{2}-7\rho _{1}\rho _{3}+26\rho _{2}\rho _{3})s^{2}}{(\rho _{1}+7\rho _{2}+19\rho _{3})^{2}R^{3}}\nonumber \\
 &  & -\frac{729(\rho _{1}^{2}+3\rho _{1}\rho _{2}-5\rho _{1}\rho _{3}+10\rho _{2}\rho _{3})s^{3}}{4(\rho _{1}+7\rho _{2}+19\rho _{3})^{2}R^{4}}+\frac{2187(\rho _{1}^{2}-2\rho _{1}\rho _{3}+2\rho _{2}\rho _{3})s^{5}}{16(\rho _{1}+7\rho _{2}+19\rho _{3})^{2}R^{6}},
\end{eqnarray}

\item \( \frac{2}{3}R\leq s\leq R: \)\begin{eqnarray}
P_{3}(s) & = & -\frac{81(9\rho _{1}\rho _{2}-9\rho _{2}^{2}+55\rho _{1}\rho _{3}-30\rho _{2}\rho _{3}-25\rho _{3}^{2})s}{8(\rho _{1}+7\rho _{2}+19\rho _{3})^{2}R^{2}}+\frac{81(9\rho _{1}\rho _{2}-\rho _{2}^{2}+19\rho _{1}\rho _{3})s^{2}}{(\rho _{1}+7\rho _{2}+19\rho _{3})^{2}R^{3}}\nonumber \\
 &  & -\frac{729(5\rho _{1}\rho _{2}-\rho _{2}^{2}+5\rho _{1}\rho _{3})s^{3}}{4(\rho _{1}+7\rho _{2}+19\rho _{3})^{2}R^{4}}+\frac{2187(2\rho _{1}-\rho _{2})\rho _{2}s^{5}}{16(\rho _{1}+7\rho _{2}+19\rho _{3})^{2}R^{6}},
\end{eqnarray}

\item \( R\leq s\leq \frac{4}{3}R: \)\begin{eqnarray}
P_{3}(s) & = & -\frac{81(64\rho _{1}-39\rho _{2}-25\rho _{3})\rho _{3}s}{8(\rho _{1}+7\rho _{2}+19\rho _{3})^{2}R^{2}}+\frac{81(8\rho _{2}^{2}+28\rho _{1}\rho _{3}-9\rho _{2}\rho _{3})s^{2}}{(\rho _{1}+7\rho _{2}+19\rho _{3})^{2}R^{3}}\nonumber \\
 &  & -\frac{729(4\rho _{2}^{2}+10\rho _{1}\rho _{3}-5\rho _{2}\rho _{3})s^{3}}{4(\rho _{1}+7\rho _{2}+19\rho _{3})^{2}R^{4}}+\frac{2187(\rho _{2}^{2}+2\rho _{1}\rho _{3}-2\rho _{2}\rho _{3})s^{5}}{16(\rho _{1}+7\rho _{2}+19\rho _{3})^{2}R^{6}},
\end{eqnarray}

\item \( \frac{4}{3}R\leq s\leq \frac{5}{3}R: \)\begin{eqnarray}
P_{3}(s) & = & -\frac{2025(\rho _{2}-\rho _{3})\rho _{3}s}{8(\rho _{1}+7\rho _{2}+19\rho _{3})^{2}R^{2}}+\frac{81(35\rho _{2}-8\rho _{3})\rho _{3}s^{2}}{(\rho _{1}+7\rho _{2}+19\rho _{3})^{2}R^{3}}\nonumber \\
 &  & -\frac{729(13\rho _{2}-4\rho _{3})\rho _{3}s^{3}}{4(\rho _{1}+7\rho _{2}+19\rho _{3})^{2}R^{4}}+\frac{2187(2\rho _{2}-\rho _{3})\rho _{3}s^{5}}{16(\rho _{1}+7\rho _{2}+19\rho _{3})^{2}R^{6}},
\end{eqnarray}

\item \( \frac{5}{3}R\leq s\leq 2R: \)\begin{equation}
\label{eq_3_shell_006}
P_{3}(s)=\frac{2187\rho ^{2}_{3}s^{2}}{(\rho _{1}+7\rho _{2}+19\rho _{3})^{2}R^{3}}-\frac{6561\rho ^{2}_{3}s^{3}}{4(\rho _{1}+7\rho _{2}+19\rho _{3})^{2}R^{4}}+\frac{2187\rho ^{2}_{3}s^{5}}{16(\rho _{1}+7\rho _{2}+19\rho _{3})^{2}R^{6}}.
\end{equation}
 
\end{enumerate}
As in the previous case, the various functional forms for \( P_{3}(s) \) are
continuous across the boundaries separating the regions. 

The \( 3 \)-shell model is of interest since actual models of neutron stars
often invoke a \( 3 \)-shell picture~\cite{Pines,Shapiro}. We note to start
with that although our \( 3 \)-shell model assumes that all shells have the
same thickness (shell radii of \( r \), \( 2r \), and \( 3r \)), we can relax
this assumption by substituting \( r\rightarrow r_{1} \), \( 2r\rightarrow r_{2} \),
and \( 3r\rightarrow r_{3} \), where \( r_{1} \), \( r_{2} \), and \( r_{3} \)
are arbitrary numbers. Similarly, the densities \( \rho _{1} \), \( \rho _{2} \),
and \( \rho _{3} \) can also assume arbitrary values, so that the results of
Eqs.~(\ref{eq_3_shell_001})--(\ref{eq_3_shell_006}) can be applied to any
realistic neutron star model~\cite{sjtu}. 

The preceding formalism can be extended to any number of shells. For a \( 4 \)-shell
model define \( \rho =\rho _{1} \) for \( 0\leq r\leq R/4 \), \( \rho =\rho _{2} \)
for \( R/4\leq r\leq R/2 \) , \( \rho =\rho _{3} \) for \( R/2\leq r\leq 3R/4 \),
and \( \rho =\rho _{4} \) for \( 3R/4\leq r\leq R \). Here \( \rho _{1} \),
\( \rho _{2} \), \( \rho _{3} \), and \( \rho _{4} \) are constants, and
\( r \) is measured from the center of the neutron star. The PDF has \( 8 \)
different functional forms specified by \( 8 \) regions:

\begin{enumerate}
\item \( R\leq s\leq \frac{1}{4}R: \)\begin{eqnarray}
P_{3}(s) & = & \frac{192(\rho _{1}^{2}+7\rho _{2}^{2}+19\rho _{3}^{2}+37\rho _{4}^{2})s^{2}}{(\rho _{1}+7\rho _{2}+19\rho _{3}+37\rho _{4})^{2}R^{3}}\nonumber \\
 &  & -\frac{576(\rho _{1}^{2}-2\rho _{1}\rho _{2}+5\rho _{2}^{2}-8\rho _{2}\rho _{3}+13\rho _{3}^{2}-18\rho _{3}\rho _{4}+25\rho _{4}^{2})s^{3}}{(\rho _{1}+7\rho _{2}+19\rho _{3}+37\rho _{4})^{2}R^{4}}\nonumber \\
 &  & +\frac{768(\rho _{1}^{2}-2\rho _{1}\rho _{2}+2\rho _{2}^{2}-2\rho _{2}\rho _{3}+2\rho _{3}^{2}-2\rho _{3}\rho _{4}+2\rho _{4}^{2})s^{5}}{(\rho _{1}+7\rho _{2}+19\rho _{3}+37\rho _{4})^{2}R^{6}},
\end{eqnarray}

\item \( \frac{1}{4}R\leq s\leq \frac{1}{2}R: \)\begin{eqnarray}
P_{3}(s) & = & -\frac{18(9\rho _{1}\rho _{2}-9\rho _{2}^{2}-9\rho _{1}\rho _{3}+34\rho _{2}\rho _{3}-25\rho _{3}^{2}-25\rho _{2}\rho _{4}+74\rho _{3}\rho _{4}-49\rho _{4}^{2})s}{(\rho _{1}+7\rho _{2}+19\rho _{3}+37\rho _{4})^{2}R^{2}}\nonumber \\
 &  & +\frac{192(\rho _{1}^{2}+7\rho _{1}\rho _{2}-7\rho _{1}\rho _{3}+26\rho _{2}\rho _{3}-19\rho _{2}\rho _{4}+56\rho _{3}\rho _{4})s^{2}}{(\rho _{1}+7\rho _{2}+19\rho _{3}+37\rho _{4})^{2}R^{3}}\nonumber \\
 &  & -\frac{576(\rho _{1}^{2}+3\rho _{1}\rho _{2}-5\rho _{1}\rho _{3}+10\rho _{2}\rho _{3}-13\rho _{2}\rho _{4}+20\rho _{3}\rho _{4})s^{3}}{(\rho _{1}+7\rho _{2}+19\rho _{3}+37\rho _{4})^{2}R^{4}}\nonumber \\
 &  & +\frac{768(\rho _{1}^{2}-2\rho _{1}\rho _{3}+2\rho _{2}\rho _{3}-2\rho _{2}\rho _{4}+2\rho _{3}\rho _{4})s^{5}}{(\rho _{1}+7\rho _{2}+19\rho _{3}+37\rho _{4})^{2}R^{6}},
\end{eqnarray}

\item \( \frac{1}{2}R\leq s\leq \frac{3}{4}R: \)\begin{eqnarray}
P_{3}(s) & = & \frac{18(9\rho _{2}^{2}-9\rho _{1}\rho _{2}-55\rho _{1}\rho _{3}+30\rho _{2}\rho _{3}+25\rho _{3}^{2})s}{(\rho _{1}+7\rho _{2}+19\rho _{3}+37\rho _{4})^{2}R^{2}}\nonumber \\
 &  & +\frac{18(64\rho _{1}\rho _{4}-183\rho _{2}\rho _{4}+70\rho _{3}\rho _{4}+49\rho _{4}^{2})s}{(\rho _{1}+7\rho _{2}+19\rho _{3}+37\rho _{4})^{2}R^{2}}\nonumber \\
 &  & +\frac{192(9\rho _{1}\rho _{2}-\rho _{2}^{2}+19\rho _{1}\rho _{3}-26\rho _{1}\rho _{4}+63\rho _{2}\rho _{4})s^{2}}{(\rho _{1}+7\rho _{2}+19\rho _{3}+37\rho _{4})^{2}R^{3}}\nonumber \\
 &  & -\frac{576(5\rho _{1}\rho _{2}-\rho _{2}^{2}+5\rho _{1}\rho _{3}-10\rho _{1}\rho _{4}+17\rho _{2}\rho _{4})s^{3}}{(\rho _{1}+7\rho _{2}+19\rho _{3}+37\rho _{4})^{2}R^{4}}\nonumber \\
 &  & +\frac{768(2\rho _{1}\rho _{3}-\rho ^{2}_{2}-2\rho _{1}\rho _{4}+2\rho _{2}\rho _{4})s^{5}}{(\rho _{1}+7\rho _{2}+19\rho _{3}+37\rho _{4})^{2}R^{6}},
\end{eqnarray}

\item \( \frac{3}{4}R\leq s\leq R: \)\begin{eqnarray}
P_{3}(s) & = & -\frac{18(64\rho _{1}\rho _{3}-39\rho _{2}\rho _{3}-25\rho _{3}^{2}+161\rho _{1}\rho _{4}-42\rho _{2}\rho _{4}-70\rho _{3}\rho _{4}-49\rho _{4}^{2})s}{(\rho _{1}+7\rho _{2}+19\rho _{3}+37\rho _{4})^{2}R^{2}}\nonumber \\
 &  & +\frac{192(8\rho _{2}^{2}+28\rho _{1}\rho _{3}-9\rho _{2}\rho _{3}+37\rho _{1}\rho _{4})s^{2}}{(\rho _{1}+7\rho _{2}+19\rho _{3}+37\rho _{4})^{2}R^{3}}\nonumber \\
 &  & -\frac{576(4\rho _{2}^{2}+10\rho _{1}\rho _{3}-5\rho _{2}\rho _{3}+7\rho _{1}\rho _{4})s^{3}}{(\rho _{1}+7\rho _{2}+19\rho _{3}+37\rho _{4})^{2}R^{4}}\nonumber \\
 &  & +\frac{768(\rho _{2}^{2}+2\rho _{1}\rho _{3}-2\rho _{2}\rho _{3})s^{5}}{(\rho _{1}+7\rho _{2}+19\rho _{3}+37\rho _{4})^{2}R^{6}},
\end{eqnarray}

\item \( R\leq s\leq \frac{5}{4}R: \)\begin{eqnarray}
P_{3}(s) & = & -\frac{18(25\rho _{2}\rho _{3}-25\rho _{3}^{2}+225\rho _{1}\rho _{4}-106\rho _{2}\rho _{4}-70\rho _{3}\rho _{4}-49\rho _{4}^{2})s}{(\rho _{1}+7\rho _{2}+19\rho _{3}+37\rho _{4})^{2}R^{2}}\nonumber \\
 &  & +\frac{192(35\rho _{2}\rho _{3}-8\rho _{3}^{2}+65\rho _{1}\rho _{4}-28\rho _{2}\rho _{4})s^{2}}{(\rho _{1}+7\rho _{2}+19\rho _{3}+37\rho _{4})^{2}R^{3}}\nonumber \\
 &  & -\frac{576(13\rho _{2}\rho _{3}-4\rho _{3}^{2}+17\rho _{1}\rho _{4}-10\rho _{2}\rho _{4})s^{3}}{(\rho _{1}+7\rho _{2}+19\rho _{3}+37\rho _{4})^{2}R^{4}}\nonumber \\
 &  & +\frac{768(2\rho _{2}\rho _{3}-\rho _{3}^{2}+2\rho _{1}\rho _{4}-2\rho _{2}\rho _{4})s^{5}}{(\rho _{1}+7\rho _{2}+19\rho _{3}+37\rho _{4})^{2}R^{6}},
\end{eqnarray}

\item \( \frac{5}{4}R\leq s\leq \frac{3}{2}R: \)\begin{eqnarray}
P_{3}(s) & = & -\frac{18(144\rho _{2}-95\rho _{3}-49\rho _{4})\rho _{4}s}{(\rho _{1}+7\rho _{2}+19\rho _{3}+37\rho _{4})^{2}R^{2}}+\frac{192(27\rho _{3}^{2}+72\rho _{2}\rho _{4}-35\rho _{3}\rho _{4})s^{2}}{(\rho _{1}+7\rho _{2}+19\rho _{3}+37\rho _{4})^{2}R^{3}}\nonumber \\
 &  & -\frac{576(9\rho _{3}^{2}+20\rho _{2}\rho _{4}-13\rho _{3}\rho _{4})s^{3}}{(\rho _{1}+7\rho _{2}+19\rho _{3}+37\rho _{4})^{2}R^{4}}+\frac{768(\rho _{3}^{2}+2\rho _{2}\rho _{4}-2\rho _{3}\rho _{4})s^{5}}{(\rho _{1}+7\rho _{2}+19\rho _{3}+37\rho _{4})^{2}R^{6}},
\end{eqnarray}

\item \( \frac{3}{2}R\leq s\leq \frac{7}{4}R: \)\begin{eqnarray}
P_{3}(s) & = & -\frac{882(\rho _{3}-\rho _{4})\rho _{4}s}{(\rho _{1}+7\rho _{2}+19\rho _{3}+37\rho _{4})^{2}R^{2}}+\frac{192(91\rho _{3}-27\rho _{4})\rho _{4}s^{2}}{(\rho _{1}+7\rho _{2}+19\rho _{3}+37\rho _{4})^{2}R^{3}}\nonumber \\
 &  & -\frac{576(25\rho _{3}-9\rho _{4})\rho _{4}s^{3}}{(\rho _{1}+7\rho _{2}+19\rho _{3}+37\rho _{4})^{2}R^{4}}+\frac{768(2\rho _{2}-\rho _{4})\rho _{4}s^{5}}{(\rho _{1}+7\rho _{2}+19\rho _{3}+37\rho _{4})^{2}R^{6}},
\end{eqnarray}

\item \( \frac{7}{4}R\leq s\leq 2R: \)\begin{eqnarray}
P_{3}(s) & = & \frac{12288\rho ^{2}_{4}s^{2}}{(\rho _{1}+7\rho _{2}+19\rho _{3}+37\rho _{4})^{2}R^{3}}-\frac{9216\rho ^{2}_{4}s^{3}}{(\rho _{1}+7\rho _{2}+19\rho _{3}+37\rho _{4})^{2}R^{4}}\nonumber \\
 &  & +\frac{768\rho ^{2}_{4}s^{5}}{(\rho _{1}+7\rho _{2}+19\rho _{3}+37\rho _{4})^{2}R^{6}}.
\end{eqnarray}
 
\end{enumerate}
A statistically tested random number generator and a special simulation technique~\cite{Seaman}
have been employed to numerically simulate the probability density functions,
and the numerical results agree with our analytical results as discussed in
Refs~\cite{sjtu,sjtu2}.

\subsection{Geometric Probability Constants}

We can also apply the preceding geometric probability techniques to evaluate
the expectation values of \( \vec{r}_{12}\cdot \vec{r}_{23} \) in \( n \)
dimensions, where \( \vec{r}_{12}=\vec{r}_{2}-\vec{r}_{1} \), \( \vec{r}_{23}=\vec{r}_{3}-\vec{r}_{2} \),
and \( \vec{r}_{1} \), \( \vec{r}_{2} \) and \( \vec{r}_{3} \) are three
independent points produced randomly inside an \( n \)-dimensional uniform
sphere of radius \( R \). The quantity \( \left\langle \vec{r}_{12}\cdot \vec{r}_{23}\right\rangle _{n} \)
is one of the geometric probability constants of interest and has important
applications in physics~\cite{sjtu,Fischbach}. In \( 3 \) dimensions,\begin{equation}
\left\langle \vec{r}_{12}\cdot \vec{r}_{23}\right\rangle _{3}=\left\langle \left( x_{2}-x_{1}\right) \left( x_{3}-x_{2}\right) +\left( y_{2}-y_{1}\right) \left( y_{3}-y_{2}\right) +\left( z_{2}-z_{1}\right) \left( z_{3}-z_{2}\right) \right\rangle .
\end{equation}
 Following Refs~\cite{sjtu,sjtu2},\begin{equation}
\label{eq_gc_001}
\left\langle \vec{r}_{12}\cdot \vec{r}_{23}\right\rangle _{n}=-n\frac{\int _{-R}^{R}x^{2}V\left( n-1,\sqrt{R^{2}-x^{2}}\right) dx}{V\left( n,R\right) }=-\frac{n}{n+2}R^{2},
\end{equation}
 where \begin{equation}
V\left( n,R\right) =\frac{\pi ^{\frac{n}{2}}R^{n}}{\Gamma \left( \frac{n}{2}+1\right) }.
\end{equation}
 We can verify Eq.~(\ref{eq_gc_001}) by applying the geometric probability
techniques directly so that \begin{equation}
\left\langle \vec{r}_{12}\cdot \vec{r}_{23}\right\rangle _{n}=-\frac{1}{2}\int _{0}^{2R}s^{2}P_{n}(s)ds=-\frac{n}{n+2}R^{2}.
\end{equation}
 Following is a short list of \( \left\langle \vec{r}_{12}\cdot \vec{r}_{23}\right\rangle _{n} \)
for selected values of \( n \):\begin{eqnarray}
\left\langle \vec{r}_{12}\cdot \vec{r}_{23}\right\rangle _{1} & = & -\frac{1}{3}R^{2},\\
\left\langle \vec{r}_{12}\cdot \vec{r}_{23}\right\rangle _{2} & = & -\frac{1}{2}R^{2},\\
\left\langle \vec{r}_{12}\cdot \vec{r}_{23}\right\rangle _{3} & = & -\frac{3}{5}R^{2},\\
\left\langle \vec{r}_{12}\cdot \vec{r}_{23}\right\rangle _{4} & = & -\frac{2}{3}R^{2},\\
\left\langle \vec{r}_{12}\cdot \vec{r}_{23}\right\rangle _{5} & = & -\frac{5}{7}R^{2}.
\end{eqnarray}
 Notice that as \( n \) becomes large \( \left\langle \vec{r}_{12}\cdot \vec{r}_{23}\right\rangle _{n}\rightarrow -R^{2} \).
If we evaluate \( \left\langle \vec{r}_{12}\cdot \vec{r}_{23}\right\rangle _{n} \)
in an \( n \)-dimensional spherical space with a Gaussian density distribution
as defined in Eq.~(\ref{eq_n_gaussian_density}) and radius \( R\rightarrow \infty  \),
then\begin{equation}
\left\langle \vec{r}_{12}\cdot \vec{r}_{23}\right\rangle _{n}=-n\sigma ^{2}.
\end{equation}
 An interesting application of our formalism in biomedical physics is the calculation
of the dynamic probability density function \( P(s,\, t) \) in \( 3 \) dimensions,
as discussed in Ref.~\cite{sjtu2}.

\section{Conclusions}

A new formalism has been presented in this paper for evaluating the analytical
probability density function of distance between two randomly sampled points
in an \( n \)-dimensional sphere with an arbitrary density. We illustrate the
power of this new geometric probability technique by demonstrating how \( n \)-dimensional
integrals can be reduced \( 1 \)-dimensional integrals, even in the presence
of \( r_{c} \). We have shown that the classical result of Eq.~(\ref{eq_kendall_santalo})
can be rederived from a simple intuitive picture, and that it can also be extended
to an \( n \)-dimensional sphere with a non-uniform density distribution. Several
density examples (Eqs.~(\ref{eq_example_1}), (\ref{eq_example_2}), (\ref{eq_example_density_2d}),
(\ref{eq_example_density_3d}), and (\ref{eq_example_density_4d})) were presented,
and the analytical results were shown to be in agreement with Monte Carlo simulations.
Applications to a variety of physical systems, such as neutron star models,
have also been discussed.

\section{Acknowledgments}

The authors wish to thank A. W. Overhauser, Michelle Parry, David Schleef, and
Christopher Tong for helpful discussions. One of the authors (S. J. T.) also
wishes to thank the Purdue University Computing Center for computing support.
Conversations on numerical algorithms with Dave Seaman and Chinh Le are also
acknowledged. This work was supported in part by the U.S. Department of Energy
under Contract No. DE-AC02-76ER01428. 

\appendix

\section{Geometry of Intersecting Circles}
\label{appendix_geometry}

\begin{figure}
\phantom{aaa}
\vfill
{\par\centering \resizebox*{6.25in}{!}{\includegraphics{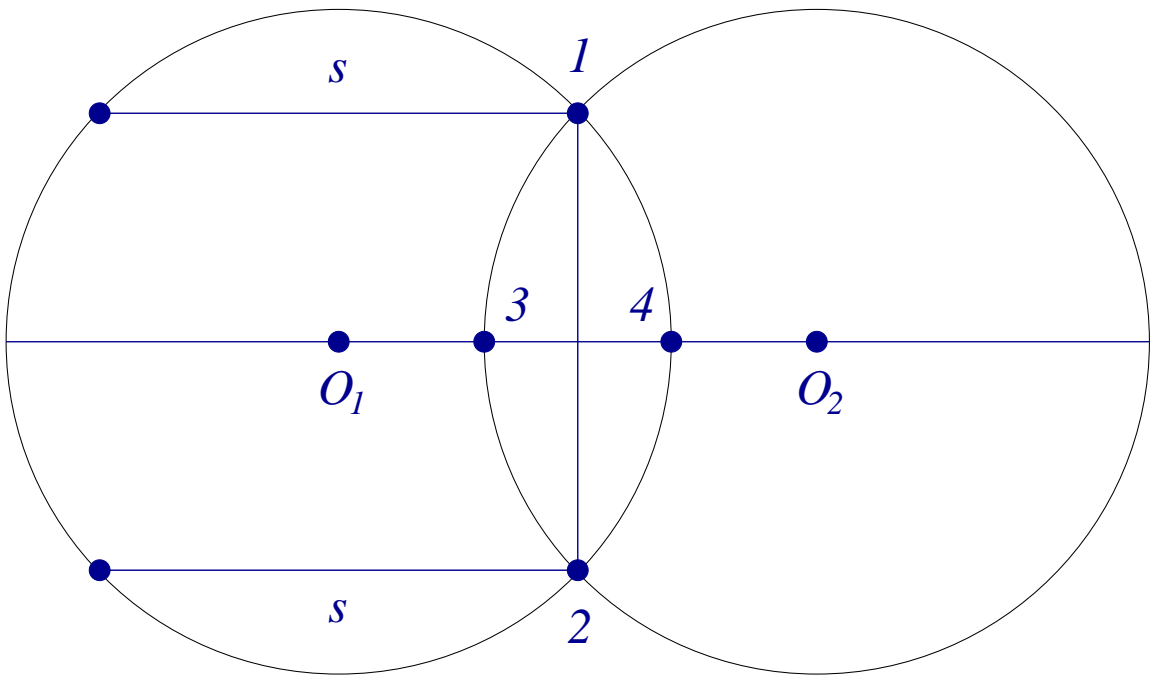}} \par}
\caption{A simple diagram showing circle \protect\( O_{1}\protect \) and \protect\( O_{2}\protect \)
are identical with the center of \protect\( O_{1}\protect \) located at \protect\( (0,0)\protect \)
and \protect\( O_{2}\protect \) at \protect\( (s,0)\protect \). \label{fig_appendix_1}}
\vfill\phantom{aaa}\end{figure}

Here we briefly show why circles \( O_{1} \) and \( O_{2} \) are identical
with the center of \( O_{1} \) located at \( (0,0) \) and \( O_{2} \) located
at \( (s,0) \) as shown in Fig.~\ref{fig_appendix_1}. Following the discussion
in Sec.~\ref{section_theory_uniform_circle}, define a Cartesian coordinate
system for points 1, 2, 3, and 4 as \begin{eqnarray}
\left( x_{1},y_{1}\right)  & = & \left( s/2,\, \sqrt{R^{2}-s^{2}/4}\right) ,\label{eq_appendix_1_a} \\
\left( x_{2},y_{2}\right)  & = & \left( s/2,\, -\sqrt{R^{2}-s^{2}/4}\right) ,\label{eq_appendix_1_b} \\
\left( x_{3},y_{3}\right)  & = & \left( s-R,\, 0\right) ,\label{eq_appendix_1_c} \\
\left( x_{4},y_{4}\right)  & = & \left( R,\, 0\right) .\label{eq_appendix_1_d} 
\end{eqnarray}
 Assume that the equation for the circle \( O_{2} \) is \begin{equation}
\label{eq_appendix_1}
(x-\alpha )^{2}+y^{2}=r^{2},
\end{equation}
 where \( (\alpha ,0) \) is the center and \( r \) is the radius. Inserting
Eqs.~(\ref{eq_appendix_1_a}), (\ref{eq_appendix_1_b}), and (\ref{eq_appendix_1_c})
into Eq.~(\ref{eq_appendix_1}) which expresses the fact that the circle \( O_{2} \)
contains points 1, 2, and 3. If \( s\neq 2R \), then the only solution is \( \alpha =s \)
and \( r=R \). This result means that circles \( O_{1} \) and \( O_{2} \)
have identical radii and the center of \( O_{2} \) is located at \( (s,0) \).

\section{Recursion Relations of $\bbox{P_{\lowercase{n}}({\lowercase{s}})}$}
\label{appendix_002}

We present in this Appendix some recursion relations for the probability density
functions \( P_{n}(s) \) which follow from the results in the text. 

\begin{equation}
P_{n}'(s)=\frac{n-1}{s}P_{n}(s)-\frac{n}{B\left( \frac{n}{2}+\frac{1}{2},\frac{1}{2}\right) }\frac{s^{n-1}}{R^{n+1}}\left( 1-\frac{s^{2}}{4R^{2}}\right) ^{\frac{n-1}{2}}.
\end{equation}
 \begin{equation}
P_{n}''(s)=-\frac{n(n-1)}{s^{2}}P_{n}(s)+\frac{2(n-1)}{s}P_{n}'(s)+\frac{n(n-1)}{4B\left( \frac{n}{2}+\frac{1}{2},\frac{1}{2}\right) }\frac{s^{n}}{R^{n+3}}\left( 1-\frac{s^{2}}{4R^{2}}\right) ^{\frac{n-3}{2}}.
\end{equation}
 \begin{equation}
\left( 1-\frac{s^{2}}{4R^{2}}\right) P_{n}''(s)-\frac{n-1}{s}\left( 2-\frac{3}{4}\frac{s^{2}}{R^{2}}\right) P_{n}'(s)+\frac{n-1}{s^{2}}\left[ n-\left( \frac{2n-1}{4}\right) \frac{s^{2}}{R^{2}}\right] P_{n}(s)=0.
\end{equation}
 \begin{eqnarray}
 & \left( 1-\frac{s^{2}}{4R^{2}}\right) \left[ P_{n}''(s)-\frac{2(n-1)}{s}P_{n}'(s)+\frac{n(n-1)}{s^{2}}P_{n}(s)\right] - & \nonumber \\
 & (n-1)\left( \frac{s^{2}}{4R^{2}}\right) \left[ \frac{1}{s}P_{n}'(s)+\frac{n-1}{s^{2}}P_{n}(s)\right] =0. & 
\end{eqnarray}
 \begin{equation}
P_{n+2}(s)=\frac{n+2}{n}\frac{s^{2}}{R^{2}}P_{n}(s)-\frac{1}{\pi }\frac{(n+2)!!}{(n+1)!!}\frac{s^{n+2}}{R^{n+3}}\left( 1-\frac{s^{2}}{4R^{2}}\right) ^{\frac{n+1}{2}}\, \, \, (n=\mathrm{even}).
\end{equation}
 \begin{eqnarray}
2P_{n}(s) & = & \frac{n}{n+2}\frac{R^{2}}{s^{2}}P_{n+2}(s)+\frac{n}{n-2}\frac{s^{2}}{R^{2}}P_{n-2}(s)\nonumber \\
 &  & -\frac{1}{\pi }\frac{n!!}{(n+1)!!}\frac{s^{n}}{R^{n+1}}\left( 1+n\frac{s^{2}}{4R^{2}}\right) \left( 1-\frac{s^{2}}{4R^{2}}\right) ^{\frac{n-1}{2}}\, \, \, (n=\mathrm{even}).
\end{eqnarray}
 \begin{equation}
P_{n+2}(s)=\frac{n+2}{n}\frac{s^{2}}{R^{2}}P_{n}(s)-\frac{1}{2}\frac{(n+2)!!}{(n+1)!!}\frac{s^{n+2}}{R^{n+3}}\left( 1-\frac{s^{2}}{4R^{2}}\right) ^{\frac{n+1}{2}}\, \, \, (n=\mathrm{odd}).
\end{equation}
 \begin{eqnarray}
2P_{n}(s) & = & \frac{n}{n+2}\frac{R^{2}}{s^{2}}P_{n+2}(s)+\frac{n}{n-2}\frac{s^{2}}{R^{2}}P_{n-2}(s)\nonumber \\
 &  & -\frac{1}{2}\frac{n!!}{(n+1)!!}\frac{s^{n}}{R^{n+1}}\left( 1+n\frac{s^{2}}{4R^{2}}\right) \left( 1-\frac{s^{2}}{4R^{2}}\right) ^{\frac{n-1}{2}}\, \, \, (n=\mathrm{odd}).
\end{eqnarray}
 \begin{equation}
P_{n+2}(s)=\frac{n+2}{n}\frac{s^{2}}{R^{2}}P_{n}(s)-\frac{1}{B\left( \frac{n}{2}+\frac{3}{2},\frac{1}{2}\right) }\frac{s^{n+2}}{R^{n+3}}\left( 1-\frac{s^{2}}{4R^{2}}\right) ^{\frac{n+1}{2}}.
\end{equation}
 \begin{eqnarray}
2P_{n}(s) & = & \frac{n}{n+2}\frac{R^{2}}{s^{2}}P_{n+2}(s)+\frac{n}{n-2}\frac{s^{2}}{R^{2}}P_{n-2}(s)\nonumber \\
 &  & -\frac{1}{n+2}\frac{1}{B\left( \frac{n}{2}+\frac{3}{2},\frac{1}{2}\right) }\frac{s^{n}}{R^{n+1}}\left( 1+n\frac{s^{2}}{4R^{2}}\right) \left( 1-\frac{s^{2}}{4R^{2}}\right) ^{\frac{n-1}{2}}.
\end{eqnarray}

\newpage

\newpage

\newpage

\newpage

\newpage

\newpage

\newpage

\newpage

\newpage

\end{document}